\documentclass[a4paper,11pt]{article}

\usepackage{jcappub} 
\usepackage[utf8]{inputenc}
\usepackage[T1]{fontenc}
\usepackage{amsmath}
\usepackage{amsfonts}
\usepackage{amssymb}
\usepackage{graphicx}
\usepackage{subcaption}

\usepackage{array}
\newcolumntype{C}{>{\centering\arraybackslash}X}

\newcommand{\nn}{\nonumber}
\newcommand{\llangle}{\left\langle}
\newcommand{\rrangle}{\right\rangle}

\newcommand{\bea}{\begin{eqnarray}}
\newcommand{\eea}{\end{eqnarray}} 
\newcommand{\beq}{\begin{equation}}
\newcommand{\eeq}{\end{equation}}

\usepackage{textcomp}
\newcommand{\degree}{\textdegree}
\usepackage{siunitx}

\begin{document}

\title{\boldmath Study the nature of dynamical dark energy by measuring the CMB polarization rotation angle} 

\author[a,1]{Hua Zhai,\note{Corresponding author.}}
\author[a,1]{Si-Yu Li}
\author[a]{Yang Liu,}
\author[b]{Yiwei Zhong,}
\author[a]{Hong Li,}
\author[a]{Yaqiong Li,}
\author[a]{Congzhan Liu,}
\author[c,d]{Mingzhe Li,}
\author[e,f]{and Xinmin Zhang}

\affiliation[a]{Key Laboratory of Particle Astrophysics, Institute of High Energy Physics, Chinese Academy of Sciences, Beijing 100049, China}
\affiliation[b]{Shing-Tung Yau Center and School of Physics, Southeast University, Nanjing 210096, People’s Republic of China}
\affiliation[c]{Interdisciplinary Center for Theoretical Study, University of Science and Technology of China, Hefei, Anhui 230026, China}
\affiliation[d]{Peng Huanwu Center for Fundamental Theory, Hefei, Anhui 230026, China}
\affiliation[e]{Theoretical physics division of Institute of High Energy Physics, Chinese Academy of Sciences, Beijing 100049, China}
\affiliation[f]{School of Nuclear Science and Technology, University of Chinese Academy of Sciences, Beijing 101408, China}
 
\emailAdd{zhaihua@ihep.ac.cn}
\emailAdd{lisy@ihep.ac.cn}
\emailAdd{liuy92@ihep.ac.cn}
\emailAdd{zhongyiwei@seu.edu.cn}
\emailAdd{hongli@ihep.ac.cn}
\emailAdd{liyq@ihep.ac.cn}
\emailAdd{liucz@ihep.ac.cn}
\emailAdd{limz@ustc.edu.cn}
\emailAdd{xmzhang@ihep.ac.cn}


\arxivnumber{2511.04459}

\abstract{Recent results from the Dark Energy Spectroscopic Instrument (DESI) support the dynamical dark energy. Intriguingly, the data favor a transition of the dark energy equation of state across $w=-1$, a hallmark of the Quintom scenario. 
In this paper, we consider a different approach to the dynamical nature of dark energy by investigating its interaction with ordinary matters, specifically the Chern-Simons (CS) interaction with photons.
In cosmology, this interaction rotates the polarized plane of the cosmic microwave background (CMB) photons, which induces non-zero polarized TB and EB power spectra.
We forecast this measurement with the Ali CMB Polarization Telescope (AliCPT) experiment.
We take the best-fit value of the isotropic rotation angle from \textit{Planck} data as our fiducial input.
We project that {10} module-year (modyr) of observations will yield an improved detection sensitivity with a significance  $\sim 5\sigma$, given a calibration precision of $0.1\si{\degree}$ in the polarization angle and {combined with \textit{Planck} HFI data}.
We also forecast AliCPT's sensitivity to the amplitude of a scale invariant spectrum of the anisotropic polarization rotation field. 
With $50$~modyr of observations, the large-aperture configuration is expected to reach $\sigma_{A_{\mathrm{CB}}}\sim10^{-2}$, offering a sixfold improvement over the small-aperture design and enabling competitive tests of spatial fluctuations in the dark energy field.}


\maketitle
\flushbottom

\section{Introduction}

Dark energy plays a central role in modern cosmology, driving the accelerated expansion of the Universe and constituting nearly 70\% of its total energy density. 
In the last two decades, observations have shown that the equation of state of dark energy is consistent with that of Einstein's cosmological constant, corresponding to a constant $w = -1$. 
However, recent results from the Dark Energy Spectroscopic Instrument (DESI) provided compelling evidence that the dark energy component may instead be dynamical in nature~\cite{DESI:2025zgx, DESI:2025wyn}.
The results indicate a notable time evolution of its equation of state, with $w$ crossing the cosmological constant boundary of $w=-1$, exhibiting behavior characteristic of the \emph{Quintom} scenario~\cite{Feng:2004ad}.
These findings have greatly renewed theoretical and observational interest in uncovering the physical origin and dynamical properties of dark energy.

In addition to studying dark energy gravitational effects through its influence on cosmic expansion and structure formation, an alternative approach is to explore its potential interactions with ordinary matter, especially in light of the growing evidence of dynamical dark energy. We consider an effective Lagrangian to describe the interactions between a dark energy scalar $\phi$ and ordinary matter. We impose a shift symmetry~($\phi \rightarrow \phi+\mathrm{const}$) to evade the experimental constraints on the fifth force. Considering the leading order in the dark energy scalar, the effective Lagrangian can be written as follow:
\begin{equation}\label{eq:interaction}
\mathcal{L}_\text{eff}=\sum_i c_i\partial_{\mu}\phi J^\mu_i,
\end{equation}
where $J^\mu_i$ represents a current associated with Standard Model particles.

In 2001, we proposed a mechanism for baryogenesis by taking $J^\mu_i$ to be proportional to the baryon current~\cite{Li:2001st,Li:2002wd}.  In a cosmological background where $\partial_0\phi\neq0$, the interaction in Eq.~(\ref{eq:interaction}) effectively introduces a chemical potential for particles carrying baryon number, which can generate a matter-antimatter asymmetry in thermo-equilibrium, thereby offering a possible pathway to explain the origin of the cosmic baryon asymmetry, $n_B/s \sim 10^{-10}$.   

If $J^\mu_i$ is related to the Chern-Simons current of electromagnetic field, Eq.~(\ref{eq:interaction}) becomes $\partial_\mu\phi  A_\nu \tilde{F}^{\mu\nu}$, or equivalently as $-(1/2)\phi  F_{\mu\nu} \tilde{F}^{\mu\nu}$. 
This term will rotate the linear polarization plane of light, an effect known as cosmic birefringence(see Ref.~\cite{Komatsu:2022nvu} for a recent review). The rotation angle $\beta$ is proportional to the change of the scalar field along the photon propagation path, i.e., $\beta \propto (\phi_0 - \phi_{\mathrm{LSS}})$. As the oldest linearly polarized light in universe, Cosmic Microwave Background (CMB) serves as a paramount medium for detecting the Chern-Simons interaction of dark energy scalar with photons. The rotation of the CMB polarization plane induces conversion between the CMB E-modes and B-modes, thus generating TB and EB correlation power spectra~\cite{Lue:1998mq,Feng:2004mq,Feng:2006dp}. The effect of a global rotation angle $\beta$ on the CMB polarization power spectra can be expressed as~\cite{Feng:2004mq,Feng:2006dp,Xia:2009ah,Li:2009rt}:
\begin{eqnarray}
\label{eq:rotated_spectra}
C_\ell^{TE,o} &=&C_\ell^{TE}\cos(2\beta)-C_\ell^{TB}\sin(2\beta), \nn\\
C_\ell^{TB,o} &=& C_\ell^{TE}\sin(2\beta)+C_\ell^{TB}\cos(2\beta), \nn\\
C_\ell^{EE,o} &=&   C_\ell^{EE}\cos^2(2\beta)+C_\ell^{BB}\sin^2(2\beta)  - C_\ell^{EB}\sin(4\beta), \nn\\
C_\ell^{BB,o} &=&C_\ell^{BB}\cos^2(2\beta)+C_\ell^{EE}\sin^2(2\beta) +C_\ell^{EB}\sin(4\beta),\nn\\
C_\ell^{EB,o} &=& \frac{1}{2} (C_\ell^{EE}-C_\ell^{BB})\sin(4\beta) + C_\ell^{EB}\cos(4\beta).
\end{eqnarray}
where $C_\ell$ represents the power spectrum before rotation, and $C_\ell^o$ denotes the rotated power spectrum.

Eqs.~(\ref{eq:rotated_spectra}) provides the basic principle for measuring the uniform CMB polarization rotation angle. 
We performed the first measurement using WMAP and BOOMERANG data in 2006~\cite{Feng:2006dp}. Subsequently, many collaborations of CMB surveys, including QUaD\cite{QUaD:2008ado}, WMAP\cite {WMAP:2012nax}, ACTPol\cite{ACTPol:2016kmo}, SPTpol\cite{Wu:2019hek} and Planck\cite{Planck:2016soo} have done this measurement. Refs.\cite{Xia:2009ah, Zhao:2015mqa, Xia:2012ck} have combined CMB and LSS observations for the analysis.

Generally, a uniform polarization rotation will be degenerate with a global miscalibration of detector polarization angles~\cite{Keating:2012ge}. CMB experiments employed a variety of techniques to determine the absolute polarization orientation of detectors. 
Common approaches include calibration using well-characterized astrophysical sources such as Tau~A~\cite{Aumont:2018epb}, artificial far-field sources~\cite{Cornelison:2020wbj}, wire-grid calibration systems \cite{Murata:2023heo}, optical modeling~\cite{Murphy:2024fna}, and diffuse Galactic foregrounds \cite{Minami:2019ruj,Minami:2020fin}. 
Another widely used method is self-calibration, which assumes the absence of any physical polarization rotation and corrects the data by minimizing the observed $TB$ and $EB$ power spectra~\cite{Keating:2012ge}.
However this approach inherently loses possibility to detect a uniform polarization rotation. 
The analysis of \textit{Planck} data by \cite{Minami:2020odp} used Galactic foreground polarization for calibration and reported $\beta = 0.35^\circ\pm 0.14^\circ$. 
Subsequently, \cite{Eskilt:2022cff} performed a joint analysis of \textit{Planck} and WMAP data with the same method, yielding a $3.6\sigma$ detection of $\beta = 0.342^\circ{}^{+0.094^\circ}_{-0.091^\circ}$, but it remained subject to modeling dependency in the Galactic foregrounds.
A later analysis of ACT data reported a consistent result, $\beta = 0.215^\circ \pm 0.074^\circ$, showing a comparable significance~\cite{Diego-Palazuelos:2025dmh}. We noticed that there has been some works on the time dependence of $\beta$~\cite{Finelli:2008jv,Gubitosi:2014cua}, however its measurements rely on the $\ell$ dependence~\cite{Ballardini:2025apf} of $\beta$ or is estimated in pixel domain~\cite{BICEPKeck:2020hhe,SPT-3G:2022ods} which is beyond the model of Eqs.~(\ref{eq:rotated_spectra}). 

Beyond the isotropic rotation, spatial variations of the polarization angle can also occur, reflecting fluctuations of the underlying dark energy field that couple to photons~\cite{Li:2008tma}. 
In such scenarios, the rotation angle $\beta(\hat{\boldsymbol{n}})$ acquires direction dependence, producing an anisotropic polarization rotation pattern over the sky. This anisotropy can be described as a random field with angular power spectrum $C_L^{\beta\beta}$, analogous to the lensing potential power spectrum $C_L^{\phi\phi}$. 
In Refs.~\cite{Li:2013vga,Zhao:2014yna}, a non-perturbative expansion approach was employed to establish the relation between the rotated and unrotated CMB power spectra under the assumption that the external field fluctuations obey statistical isotropy.

When one considers a specific realization of the rotation pattern across the sky, the statistical isotropy of the CMB is broken, coupling off-diagonal multipoles with $\ell\neq\ell'$. 
This coupling allows the rotation field to be reconstructed through a quadratic estimator technique, in close analogy to CMB lensing reconstruction~\cite{Gluscevic:2009mm,Yadav:2009eb}. 
The first implementation of this method was presented in Ref.~\cite{Gluscevic:2009mm} using WMAP7 data.
Subsequent analyses, such as those by the POLARBEAR~\cite{POLARBEAR:2015ktq}, SPTpol~\cite{SPT:2020cxx}, ACT~\cite{Namikawa:2020ffr}, and BICEP/Keck~\cite{BICEPKeck:2022kci} collaborations, have applied similar methods, though none has yet detected a statistically significant signal. 
The best current 95\% upper bound on the amplitude of a scale-invariant rotation spectrum is $A_{\mathrm{CB}}\leq0.044$~\cite{BICEPKeck:2022kci}, defined through $L(L+1)C_L^{\beta\beta}/(2\pi)=A_{\mathrm{CB}}\times10^{-4}\text{ [rad}^2]$.

In this work, we take the Ali CMB Polarization Telescope (AliCPT) experiment~\cite{Li:2017drr,Salatino:2020skr} as an example to forecast its capability in detecting both the isotropic and anisotropic polarization rotation angles.
AliCPT, located in Tibet, China, is a high-altitude ground-based CMB mission in the Northern Hemisphere.
The available sky coverage can be up to approximately 70\% of the sky~\cite{Li:2018rwc}.
Its scientific objectives include measuring the tensor-to-scalar ratio $r$ of primordial gravitational waves and probing the dynamical nature of dark energy via the CMB polarization rotation angle. 
The first phase of AliCPT~(AliCPT-1) operates in two frequency bands, 95 GHz and 150 GHz, with a telescope aperture of 72 cm, and successfully achieved first light in early 2025.

The structure of this paper is organized as follows: Section \ref{sec:fisher_iso} introduces the methodology for measuring the uniform rotation angle and presents forecast results based on the AliCPT experiment; Section \ref{sec:forecast_aniso} conducts a preliminary forecast on the measurement of anisotropic rotation angle, especially considering a planned large aperture telescope. Section \ref{sec:summary} is our conclusion.

\section{Forecast for Isotropic Polarization Rotation}\label{sec:fisher_iso}

\subsection{Methodology}\label{sec:Methodology}

To estimate the Chern-Simons interaction-induced rotation angle $\beta$ in CMB polarization, we employ the so-called Minami-Komatsu method that utilizes Galactic foreground polarization to break the degeneracy between $\beta$ and the instrumental polarization miscalibration angle $\alpha_i$, where $i$ labels the frequency band. The underlying principle is that the polarization orientation of CMB is influenced by ($\alpha_i$ + $\beta$), whereas the Galactic foreground radiation is affected solely by $\alpha_i$. 
Assuming that the observed microwave sky signal comprises CMB, foreground radiation, and noise, we apply Eqs.~(\ref{eq:rotated_spectra}) to both CMB and foreground, thereby eliminating the original power spectrum of the foreground. This yields a relationship that encompasses the original CMB power spectrum and the observed power spectrum:
\begin{eqnarray}
\label{eq:likelihood_eq}
C_\ell^{E^iB^j, o} 
&=&   \left(C_\ell^{E^iE^j, o}\sin(4\alpha^{j})-C_\ell^{B^iB^j, o}\sin(4\alpha^{i}) \right)\frac{1}{\cos(4\alpha^{i}) + \cos(4\alpha^{j})}  \nn\\
&&+  \frac{1}{2} \left(C_\ell^{E^iE^j, cmb, th} -  C_\ell^{B^iB^j, cmb, th}\right)  \frac{\sin(4\beta)}{\cos(2\alpha^{i}+2\alpha^{j})}\nn\\
&&-\frac{1}{2}\left(N_\ell^{E_iE_j}- N_\ell^{B_iB_j}\right)\tan(2\alpha^{i}+2\alpha^{j})+ N_\ell^{E^iB^j}\nn\\
&&+ \frac{1}{2\cos(2\alpha^i+2\alpha^j)}\left[(C_\ell^{E^iB^j, cmb}+C_\ell^{E^jB^i, cmb})\cos(4\beta) + C_\ell^{E^iB^j, fg}+C_\ell^{E^jB^i, fg}\right]\nn\\
&& +\frac{1}{2\cos(2\alpha^i-2\alpha^j)}\left[ C_\ell^{E^iB^j, cmb}-C_\ell^{E^jB^i, cmb} + C_\ell^{E^iB^j, fg}-C_\ell^{E^jB^i, fg}\right],
\end{eqnarray}
where $C_\ell^{cmb, th}$ denotes the original CMB power spectrum, while $C_\ell^{fg}$ and $N_\ell$ represent the foreground and noise power spectra, respectively. 
In this work, we use the exact trigonometric form for the isotropic rotation angle in the calculations, differing from the small-angle approximation adopted in \cite{Dou:2025luz}.

The third line of Eq.~(\ref{eq:likelihood_eq}) arises from deviations of the noise from whiteness or spatial inhomogeneity. 
The fourth and fifth lines contain $C_\ell^{EB,\mathrm{cmb}}$ and $C_\ell^{EB,\mathrm{fg}}$, which represent the intrinsic EB correlations of the CMB and foregrounds, respectively. 
{CMB polarization exhibits no intrinsic EB correlation in the standard $\Lambda$CDM model, whereas there are hints for $EB$ correlation in foreground. Planck observations~\cite{Clark:2021kze} have revealed a positive TB correlation in the dust polarization power spectrum, which may originate from magnetically misaligned filamentary structures in the neutral hydrogen component of the interstellar medium~\cite{Clark:2021kze}. The same mechanisms could also potentially induce a nonzero EB correlation. To account for the effect of foreground EB correlation in the estimation of polarization angles, we follow the approach of Ref.~\cite{Diego-Palazuelos:2022cnh}, modeling the foreground EB contribution with a template scaled by an amplitude parameter. Neglecting third line of Eq.~(\ref{eq:likelihood_eq}) and CMB $EB$ correlation,} we construct a Gaussian likelihood function for the joint estimation of $\alpha$ and $\beta$ as follows~\cite{Eskilt:2022cff}:
\begin{eqnarray}
\mathcal{L} &=& -2\ln L = \sum_\ell \left[\vec{X}_\ell^T\Xi^{-1}_\ell \vec{X}_\ell  + \ln |\Xi_\ell|\right],\nn\\
\vec{X}_\ell &=& \vec{O}_\ell + \vec{T}^c_\ell + \vec{T}^f_\ell,
\end{eqnarray}
where $\vec{O}_\ell = \left\{O_\ell^{00}, O_\ell^{01}, \dots, O_\ell^{ij}, \dots \right\}^T$ denotes the observation vector, $\vec{T}^c_\ell = \left\{ T_\ell^{c, ij}\right\}^T$ and $\vec{T}^f_\ell = \left\{  T_\ell^{f, ij} \right\}^T$ represent CMB theoretical power spectrum and foreground template vector respectively, where $i$ and $j$ are frequency band indices. $\Xi_\ell$ is the covariance matrix. For $n$ frequency bands, both $O_\ell^{ij}$ and $T^{c/f, ij}_\ell$ contain $n^2$ elements, encompassing all auto and cross power spectra. The explicit expressions for $O_\ell^{ij}$ and $T_\ell^{c/f,ij}$ are given below:
\begin{eqnarray}
O^{ij}_\ell  &=& \vec{A}^{ij, T}_o\vec{C}^{ij, o}_\ell,~~T^{c/f, ij}_\ell  =\vec{A}^{ij, T}_{c/f}\vec{C}^{ij, c/f}_\ell, \nn\\
\vec{C}^{ij, o}_\ell &= & \left\{ C^{E_iE_j, o}_\ell,C^{B_iB_j, o}_\ell,C^{E_iB_j, o}_\ell \right\}^T, \nn\\
\vec{C}_\ell^{ij, c} &=& \left\{ C^{E_iE_j, cmb}_\ell,C^{B_iB_j, cmb}_\ell \right\}^T,~~
\vec{C}_\ell^{ij, f} = \left\{ C^{E_iB_j, fg}_\ell,C^{B_iE_j, fg}_\ell \right\}^T, 
\end{eqnarray}
where the coefficient vector $\vec{A}$ are,
\begin{eqnarray}
\vec{A}_o^{ij} &=& \left\{ \frac{-\sin(4\alpha^j), \sin(4\alpha^i)}{\cos(4\alpha^i)+\cos(4\alpha^j)}, 1\right\}^T,~~~\vec{A}_c^{ij} =  - \frac{\sin(4\beta)}{2\cos(2\alpha^i+2\alpha^j)} \left\{1, -1\right\}^T, \nn\\
\vec{A}_f^{ij, T} &=& - \frac{\mathcal{A}}{2}\left\{ \sec(2\alpha^i+2\alpha^j) +\sec(2\alpha^i-2\alpha^j),\sec(2\alpha^i+2\alpha^j) -\sec(2\alpha^i-2\alpha^j)\right\}^T,
\end{eqnarray}
{here $\mathcal{A}$ is the amplitude parameter of foreground EB template}.

{Typically, $\vec{T}^{c/f}_\ell$ do not significantly contribute to the covariance matrix $\Xi$ if the CMB power spectrum and the foreground EB are known precisely. But to obtain a more realistic likelihood function, we follow the approach in Ref.~\cite{Diego-Palazuelos:2022cnh} and incorporate the variances of $\vec{T}^{f}_\ell$, as well as its covariance with $\vec{O}_\ell$, into the covariance matrix. However, unlike Ref.~\cite{Diego-Palazuelos:2022cnh}, we do not include the variance of  $\vec{T}^{c}_\ell$ or its covariance with $\vec{O}_\ell$ as these contribution are suppressed by the small $\sin^2(4\beta)$ factor, as shown in  Appendix.~\ref{append:covariancce}(This is also confirmed numerically in that section).} The covariance matrix component is written as,
\begin{eqnarray}\label{eq:cov_exp}
\Xi_\ell^{pq} &=& \mathrm{Cov}\left(O^{ij}_\ell+T^{ij, c}_\ell+T^{ij, f}_\ell , O^{i'j'}_\ell+T^{i'j', c}_\ell+ T^{i'j', f}_\ell\right) 
\nn\\	
&=& \Xi_\ell^{pq, o*o} + \Xi_\ell^{pq, o*f} + \Xi_\ell^{pq, f*o} + \Xi_\ell^{pq, f*f}
\end{eqnarray}

We assume no correlations between different $\ell$ modes. For each part, the covariance matrix elements is given by:
\begin{eqnarray}\label{eq:cov_def}
\Xi_\ell^{pq, o*o} &=& \mathrm{Cov}\left(O^{ij}_l, O^{i'j'}_l\right) 
= \vec{A}_o^{ij, T} \mathrm{Cov}\left(\vec{C}^{ij, o}_l, (\vec{C}^{i'j', o}_l )^T \right) \vec{A}_o^{i'j'} = \vec{A}_o^{ij, T} Q^{iji'j',o*o} \vec{A}^{i'j'},\nn\\
\Xi_\ell^{pq, f*f} & =&  \vec{A}_f^{ij, T} Q^{iji'j',f*f}  \vec{A}_f^{i'j'} ,   \nn\\
\Xi_\ell^{pq, o*f}  &=&  \vec{A}_o^{ij, T} Q^{iji'j',o*f}  \vec{A}_f^{i'j'} , ~~
\Xi_\ell^{pq, f*o}  = \vec{A}_f^{ij, T} Q^{iji'j',f*o}  \vec{A}_o^{i'j'} , 	
\end{eqnarray}
where $i$ and $j$ denote the frequency band combination corresponding to the $p$-th component of the observation vector, while $i'$ and $j'$ correspond to the $q$-th component. The quantity $Q^{iji'j',o*o}$ represents the covariance matrix of the observed $EE$, $BB$, and $EB$ power spectra, and is given by:
\begin{eqnarray}\label{eq:qoo}
Q^{iji'j', o*o} &=& \left( {\begin{array}{ccc}
		\mathrm{Cov}\left( C_\ell^{E^{i}E^{j}, o}, C_\ell^{E^{i'}E^{j'}, o}\right)	& \mathrm{Cov}\left( C_\ell^{E^{i}E^{j}, o}, C_\ell^{B^{i'}B^{j'}, o}\right) &
		\mathrm{Cov}\left( C_\ell^{E^{i}E^{j}, o}, C_\ell^{E^{i'}B^{j'}, o}\right) \\
		\mathrm{Cov}\left( C_\ell^{B^{i}B^{j}, o}, C_\ell^{E^{i'}E^{j'}, o}\right)	& \mathrm{Cov}\left( C_\ell^{B^{i}B^{j}, o}, C_\ell^{B^{i'}B^{j'}, o}\right) &
		\mathrm{Cov}\left( C_\ell^{B^{i}B^{j}, o}, C_\ell^{E^{i'}B^{j'}, o}\right) \\
		\mathrm{Cov}\left( C_\ell^{E^{i}B^{j}, o}, C_\ell^{E^{i'}E^{j'}, o}\right)	& \mathrm{Cov}\left( C_\ell^{E^{i}B^{j}, o}, C_\ell^{B^{i'}B^{j'}, o}\right) &
		\mathrm{Cov}\left( C_\ell^{E^{i}B^{j}, o}, C_\ell^{E^{i'}B^{j'}, o}\right) \\	
\end{array} } \right).\nn\\
\end{eqnarray}
{The explicit expressions for other listed power spectrum covariance matrix $Q^{iji'j', f*f}$, $Q^{iji'j', o*f}$, $Q^{iji'j', f*o}$ are in Appendix.~\ref{append:covariancce}}
	
The covariance of the power spectrum terms in Eqs.~(\ref{eq:qoo})can be calculated by using approximate covariance as in Ref.~\cite{Minami:2020fin}:
\begin{equation} \label{eq:cov_spectrum_law}
\mathrm{Cov}( C_\ell^{X_iY_j, o}, C_\ell^{Z_{s}W_{t}, o}) 
= \frac{1}{(2l+1)f_{sky}} \left(C_\ell^{X_iZ_{s}}C_\ell^{Y_jW_{t}}  + C_\ell^{X_iW_{t}}C_\ell^{Y_jZ_{s}} \right),
\end{equation}
where $f_{sky}$ is effective sky fraction, calculated according to Eq~.(22) in Ref.~\cite{Eskilt:2022cff}.

The likelihood function has so far been presented in terms of a continuous multipole $\ell$. In practice, the power spectra are usually binned in $\ell$ to mitigate noise. We adopt a simple top-hat binning scheme, under which the binned power spectra and their covariance matrix are computed accordingly. For example, consider the observed power spectrum vector $O^{ij}_\ell$,
\begin{eqnarray}
\hat{O}^{ij}_{b} &=& \frac{1}{\Delta l}\sum_{l\in b}\hat{O}^{ij}_l, \\
\Xi_{b, ij,i'j'} &=&  \mathrm{Cov}\left(\hat{O}^{ij}_b , \hat{O}^{i'j'}_b  \right) = \frac{1}{(\Delta l)^2} \sum_{l\in b}  \mathrm{Cov}\left(\hat{O}^{ij}_l , \hat{O}^{i'j'}_l \right).\label{eq:binned_cov} 
\end{eqnarray}
The corresponding binned likelihood is then given by:
\begin{equation}\label{eq:eb_multi_likelihood_bined}
\mathcal{L} = \sum_{b=1}^{nbins} \left[ \left(\vec{O}_b + \vec{T}^c_b + \vec{T}^f_b \right)^{\mathrm{T}}\Xi^{-1}_b\left(\vec{O}_b + \vec{T}^c_b + \vec{T}^f_b\right) + \ln |\Xi_b| \right].
\end{equation}

In a Bayesian framework, we will use calibration data for the polarization miscalibration angle~$\alpha_i$ as a prior. Such calibration data may be obtained from either astronomical or artificial polarized sources. The prior helps estimate both $\alpha$ and $\beta$ together. {Suppose that calibration information is available for $m$ frequency bands, and that for each calibrated band the prior on $\alpha_i$ is independent and follows a Gaussian distribution, $\alpha_i \sim \mathcal{N}(\bar{\alpha}_i, \sigma^{cali}_{\alpha_i})$. The likelihood function is then modified as}:
\begin{equation}\label{eq:eb_multi_likelihood_bined_prior}
\mathcal{L} = \sum_{b=1}^{nbins} \left[ \left(\vec{O}_b  + \vec{T}^c_b + \vec{T}^f_b\right)^{\mathrm{T}}\Xi^{-1}_b\left(\vec{O}_b  + \vec{T}^c_b + \vec{T}^f_b \right)+ \ln |\Xi_b|\right]+ \sum_{i=1}^m \frac{(\alpha_i-\bar{\alpha_i})^2}{(\sigma^{cali}_{\alpha_i})^2}.
\end{equation}

Since the likelihood functions in Eq.~(\ref{eq:eb_multi_likelihood_bined_prior}) have relatively simple trigonometric analytic forms, we can directly compute the corresponding Fisher information matrix.  
The Fisher matrix is derived from its definition:
\bea\label{eq:fisher_def1}
F_{\theta\phi} &=& \llangle \left(\frac{\partial}{\partial_\theta}  \mathcal{L}\right) \left(\frac{\partial }{ \partial_\phi} \mathcal{L} \right)\rrangle \nn\\
&=&  \sum_{b}\left[ \frac{1}{2}\text{Tr}[\Xi^{-1}_b (\kappa_{b\theta\phi}+\lambda_{b\theta\phi}) ] + \text{Tr}[ \Xi^{-1}_b D^L_{b,\phi} \Xi^{-1} D^L_{b,\theta} ]  - \frac{1}{2} \text{Tr}[ \Xi^{-1}_b \Xi_{b, \theta}\Xi^{-1}_b \Xi_{b, \phi} ]\right] \nn\\
&& + \frac{1}{4} \left(\sum_{b}   \mathrm{Tr}( \Xi^{-1}_b\Xi_{b,\theta}) \right) \left(\sum_{b}    \mathrm{Tr}( \Xi^{-1}_b\Xi_{b,\phi}) \right) + \frac{\Delta_{\theta\phi}}{(\sigma_\theta^{cali})^2}, \nn\\
\kappa_{pq, \theta\phi} &=&   \sum_{x,z \in [o, f]}\left[ \vec{A}^{p, T}_{x,\theta}Q^{pq, x*z}\vec{A}^{q}_{z, \phi} + \vec{A}^{p, T}_{x,\phi}Q^{pq, x*z}\vec{A}^{q}_{z, \theta}\right],\nn\\
\lambda_{pq, \theta\phi} &=&  \sum_{x \in [o, c, f]}\left( \vec{A}^{p, T}_{x,\theta} \vec{{C}}^{p,x}\right) \sum_{z \in [o, c, f]} \left( \vec{A}^{q, T}_{z,\phi}\vec{{C}}^{q,z}\right)  + (q \leftrightarrow p),\nn\\
D^L_{pq,\theta} &=&  \sum_{x,z \in [o, f]}\vec{A}^{p, T}_{x,\theta}Q^{pq, x*z}\vec{A}^{q}_{z},\label{eq:fisher_matrix_expression} 
\eea
where $b$ denotes the multipole bin index, $p, q$ are vector/matrix element indices, and $\theta,\phi \in \{\alpha, \beta\}$ label the parameters of interest, $\Delta_{\theta\phi}$ equals to $1$ when $\theta=\phi \in \{\alpha_i, i=0, ..., m\}$, otherwise equals to zero. The calibration uncertainty $\sigma^{cali}_\theta$ is defined only for the polarization miscalibration angle $\alpha$. The derivation of this Fisher matrix formula and the explicit expressions of these derivative terms are given in Appendix~\ref{append:fisher}.

\subsection{Fisher forecast configuration}
\label{sec:alicpt_config}

\begin{figure}[htbp]
	\centering    
	\includegraphics[width=0.8\columnwidth]{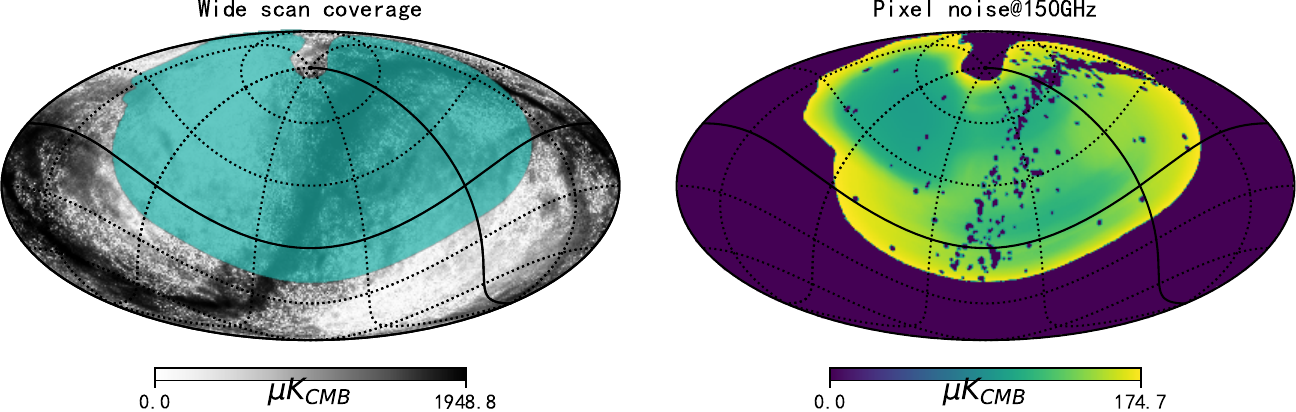}
	\caption{The left panel shows the sky coverage of the AliCPT-1 wide scan, with the background representing the dust polarization intensity from \textit{Planck} $353$GHz map. The right panel displays noise standard deviation corresponding to $1$ module year of observation in the selected {46\%} sky area.}\label{fig:wide_scan}
\end{figure}

We perform the Fisher forecast for the AliCPT experiment.
The AliCPT observatory, located at latitude $32^\circ 18' 38''$N and longitude $80^\circ 1' 50''$E at an altitude of 5,250 m, operates in two scanning modes: a deep survey covering approximately 10\% of the sky optimized for primordial gravitational wave detection, and a wide-field survey covering about 50\% of the sky, designed for CMB polarization rotation measurements.
The wide-field coverage is shown in the left panel of Fig.~\ref{fig:wide_scan}. 
To constrain the isotropic rotation angle $\beta$, we select a region with relatively uniform noise and {exclude the CO line and polarized point source with the union of point-source masks of all the \textit{Planck} maps, resulting in an effective sky fraction of about 46\% }(right panel of Fig.~\ref{fig:wide_scan}).
The observing frequency bands and nominal noise levels of AliCPT-1 are summarized in Table~\ref{tab:bands_design}.

To enhance the frequency covering range, we combine AliCPT-1 data with \textit{Planck} observations, including the LFI (30, 44, 70 GHz) and HFI (100, 143, 217, 353 GHz) channels, whose beam parameters and noise levels are taken from Table 4 of Ref.~\cite{Planck:2018nkj}.

For the fiducial cosmology, we adopt $\beta = 0.35^\circ$, the best-fit value reported in~\cite{Minami:2020odp}. Our independent fit to the \textit{Planck} 2018 polarization data using the likelihood in Section~\ref{sec:Methodology} yields a consistent result, supporting this choice. 
The cosmological parameters follow the \textit{Planck} 2018 best-fit $\Lambda$CDM model.
Instrumental noise is modeled as uniform white noise, scaling with the number of module-years $n$ as $w_p^{-1/2}/\sqrt{n}$ for AliCPT-1, and using $w_p^{-1/2}$ values from Ref.~\cite{Planck:2018nkj} for \textit{Planck} channels.
Foreground power spectra are modeled using \texttt{NaMaster}~\cite{Alonso:2018jzx} to compute binned $EE$ and $BB$ spectra (bin width $\Delta\ell = 10$) from PYSM simulations~\cite{Zonca:2021row} over the 46\% sky mask. These binned spectra are interpolated to produce smooth $\ell$-dependent functions as foreground input. {We use the medium-complexity foreground model~\cite{Pan-ExperimentGalacticScienceGroup:2025vcd} in simulation, specifically adopting the components ‘d10’, ‘s5’, ‘f1’, ‘a2’ and ‘co3’, which correspond to thermal dust, synchrotron, free-free, anomalous microwave emission (AME) and CO emission, respectively. Note that ‘a2’ is used in place of ‘a1’ for the original medium-complexity model}. Finally, CMB power spectrum is rotated by $\alpha_i+\beta$ following Eqs.~(\ref{eq:rotated_spectra}) where $\alpha_i$ are all chosen to be zero, while foreground power spectrum is rotated by $\alpha_i$. Then the sum of rotated CMB and foreground power spectrum are smoothed with beam window function, and added by the white noise to simulate the observed power spectrum. The observed power spectrum applied to Eqs.~(\ref{eq:fisher_matrix_expression}) are binned with width $\Delta\ell = 20$ as the manner of Ref.~\cite{Minami:2020odp}. 

To validate the Fisher implementation, we perform a comparison between Fisher-forecast uncertainty of polarization rotation angle and the dispersion obtained from $200$ minimum likelihood estimations in simulations. Further detail of simulations can be found in Appendix \ref{append:simulation}.

\begin{table}[ht]
	\caption{AliCPT-1 frequency band parameters and the noise level per module per observing season~\cite{Salatino:2020skr}}
	\centering
	\setlength{\tabcolsep}{4mm}{
		\begin{tabular}{c|c|c} 
			\hline  
			freq(GHz)  &  FWHM(arcmin)  &   $w_p^{-1/2}(\mu$ K-arcmin/mod/year)  \\
			\hline
			95 & 19  &     58.2  \\ 
			\hline
			150 & 11 &    87.3    \\ 
\hline	     			           		
	\end{tabular}}
	\label{tab:bands_design}
\end{table}

\subsection{Forecasting Results on polarization angle uncertainty}

\begin{figure}[htbp]
	\centering    
	\includegraphics[width=0.9\columnwidth]{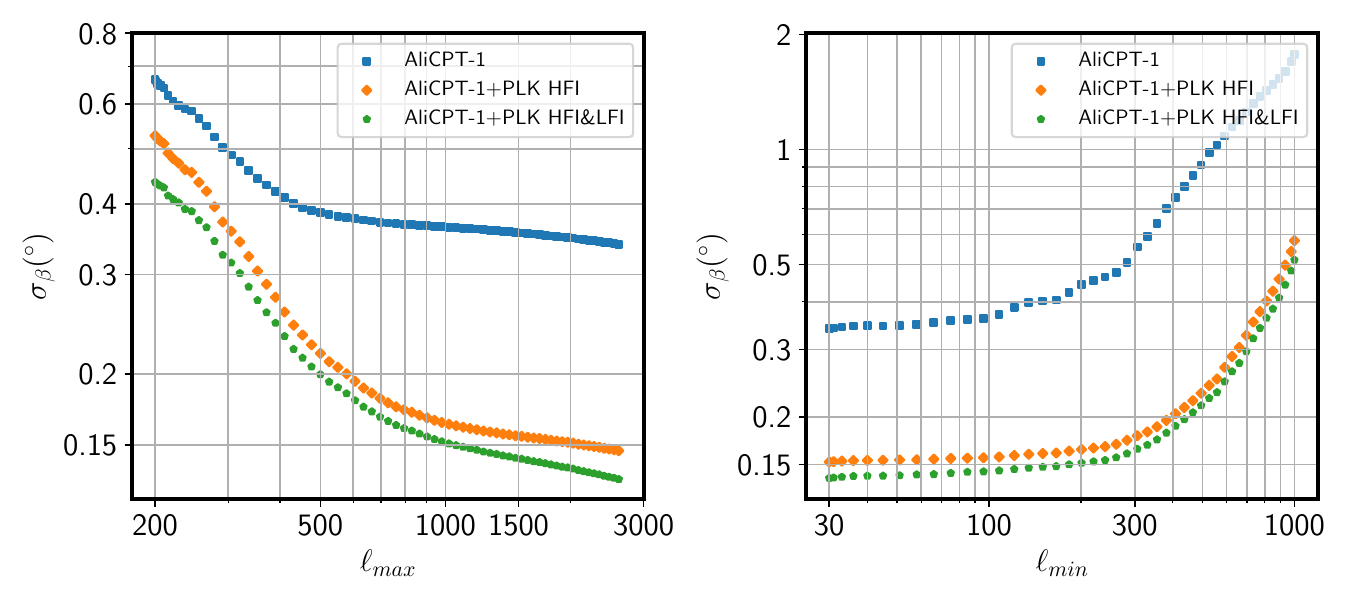}
	\caption{Constraints on $\beta$ with varied $\ell_{\max}$ ($\ell_{\min}=71$ fixed; left panel) and varied $\ell_{\min}$ ($\ell_{\max}=1500$ fixed; right panel), for AliCPT-1 (95/150 GHz, 10 module-years), PLK HFI (100, 143, 217, 353 GHz), and PLK LFI (30, 44, 70 GHz) simulations.}
	\label{fig:sig_beta_vs_ell}
\end{figure}

{In this section, we set the foreground template amplitude parameter $\mathcal{A}=0$. The impact of a nonzero $\mathcal{A}$ on $\sigma_\beta$ is investigated in Section~\ref{sec:bias_study}. In the forecast, we examine the power spectrum over  the multipoles range $\ell \in [30,3000]$}, which encompasses the angular scales where both the CMB signal and Galactic foregrounds contribute significantly.
The influence of the multipole range on the constraint of $\beta$ is shown in Fig.~\ref{fig:sig_beta_vs_ell}.
Since Galactic foregrounds dominate at low $\ell$, 
increasing $\ell_{\max}$ beyond $\sim1500$ yields little additional improvement, as high-$\ell$ modes contribute marginally to breaking the degeneracy between $\alpha$ and $\beta$. On the other hand, the constraining power quickly saturates once $\ell_{\min} \lesssim 100$, which may be due to the cosmic variance of power spectrum at low $\ell$. 
These results indicate that the effective multipole range for constraining isotropic polarization rotation angle with AliCPT-1’s wide-scan configuration is approximately $100 \lesssim \ell \lesssim 1500$. {In the remainder of this paper,and to maintain consistency with the simulation in appendix,we use the multipole range $71 \lesssim \ell \lesssim 1500$.}

The results in Fig.~\ref{fig:sig_beta_vs_ell} also demonstrate that incorporating additional frequency bands improves the estimation precision of $\beta$. In particular, combining AliCPT-1 with \textit{Planck} HFI data leads to a significant reduction in $\sigma_\beta$, while further inclusion of \textit{Planck} LFI bands yields only marginal additional improvement.

\begin{figure}[htbp]
	\centering    
	\includegraphics[width=0.65\columnwidth]{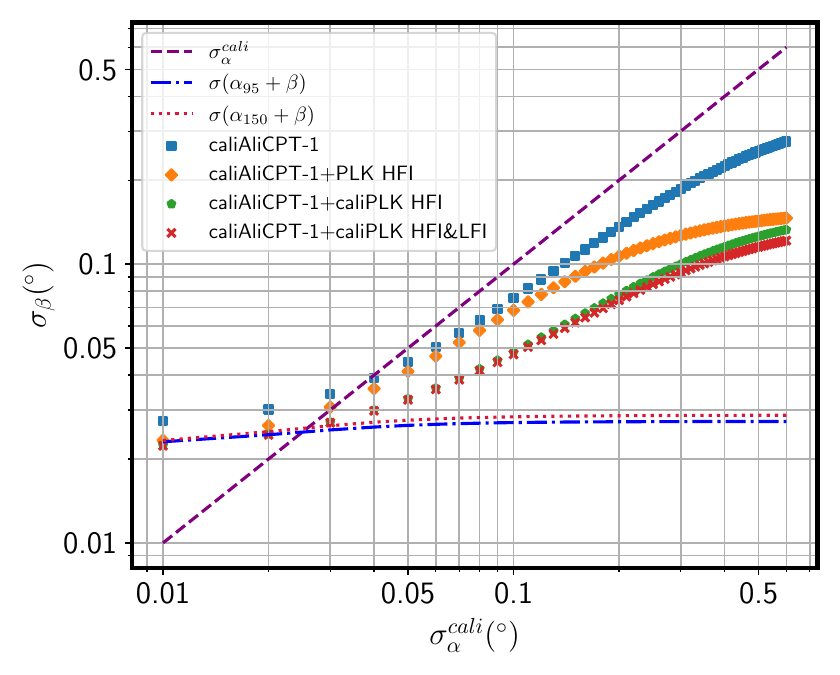}
	\caption{Evolution of $\sigma_\beta$ with $\sigma^{cali}_\alpha$ for different data combinations. "caliAliCPT-1": prior applied to both AliCPT-1 bands; "caliPLK": same prior assumed for \textit{Planck} bands.}\label{fig:sig_beta_vs_prior}
\end{figure}

We present the impact of taking into account of the calibration uncertainty of $\sigma^{cali}_\alpha$ on $\sigma_\beta$ in Fig.~\ref{fig:sig_beta_vs_prior}. 
{Recent studies using drone-carried polarized source or wire-grid methods have demonstrated promising results in reducing CMB polarization angle calibration errors to approximately $0.1\si{\degree}$~\cite{Murata:2023heo, Dunner:2024zoy}.  We take the same method in Ref.~\cite{Murata:2023heo} to calibrate the polarization angle. In forecast, we consider scenarios where the calibration uncertainty varies from $0.5\si{\degree}$ to $0.01\si{\degree}$, noting that values below $0.1\si{\degree}$ are primarily for theoretical exploration.} 
We consider three cases: AliCPT-1 with 10 module-years alone, AliCPT-1 + \textit{Planck} HFI, and AliCPT-1 + \textit{Planck} HFI/LFI.
For AliCPT-1, we assume a common polarization angle calibration uncertainty across its two frequency bands. 
The results indicate that:
\begin{itemize}
    \item When {$\sigma^{cali}_\alpha \gtrsim 0.03\si{\degree}$}, the constraint on  $\beta$ improves as the prior uncertainty tightens, and consistently remains below $\sigma^{cali}_\alpha$ itself. This indicates that the combination of foreground and external calibrator yields a more precise measurement of $\beta$ than either method could achieve independently.
    
    \item When $\sigma^{cali}_\alpha \lesssim 0.03\si{\degree}$, although $\sigma_\beta$ continues to decrease with tighter $\sigma^{cali}_\alpha$, yet it becomes larger than $\sigma^{cali}_\alpha$. This occurs because now the high precision of the calibrator surpasses the statistical power of CMB data. Consequently, $\sigma(\alpha+\beta)$ dominates the error, and its magnitude is primarily determined by CMB data amount (as indicated by the blue dashed and red dotted lines in Fig.~\ref{fig:sig_beta_vs_prior}). Here, $\sigma(\alpha+\beta)$ is calculated from the inverse of the Fisher matrix, which serves as the covariance matrix for $\alpha_i$ and $\beta$, using $\sigma(\alpha+\beta)=\sqrt{(F^{-1})_{\alpha\alpha} +(F^{-1})_{\beta\beta} +2(F^{-1})_{\alpha\beta}}$.
    
    \item  Not only the value of $\sigma^{cali}_\alpha$, but also the number of calibrated frequency bands used plays a significant role in constraining $\beta$. As depicted, when $\sigma^{cali}_\alpha$ is better than $0.1\si{\degree}$, simply adding \textit{Planck} data gives only a small improvement. However, if we also apply a calibration prior to the \textit{Planck} bands, the constraint on $\beta$ improves significantly. This shows that using more well-calibrated frequency bands is crucial.
\end{itemize}

\begin{figure}[htbp]
	\centering    
	\includegraphics[width=0.65\columnwidth]{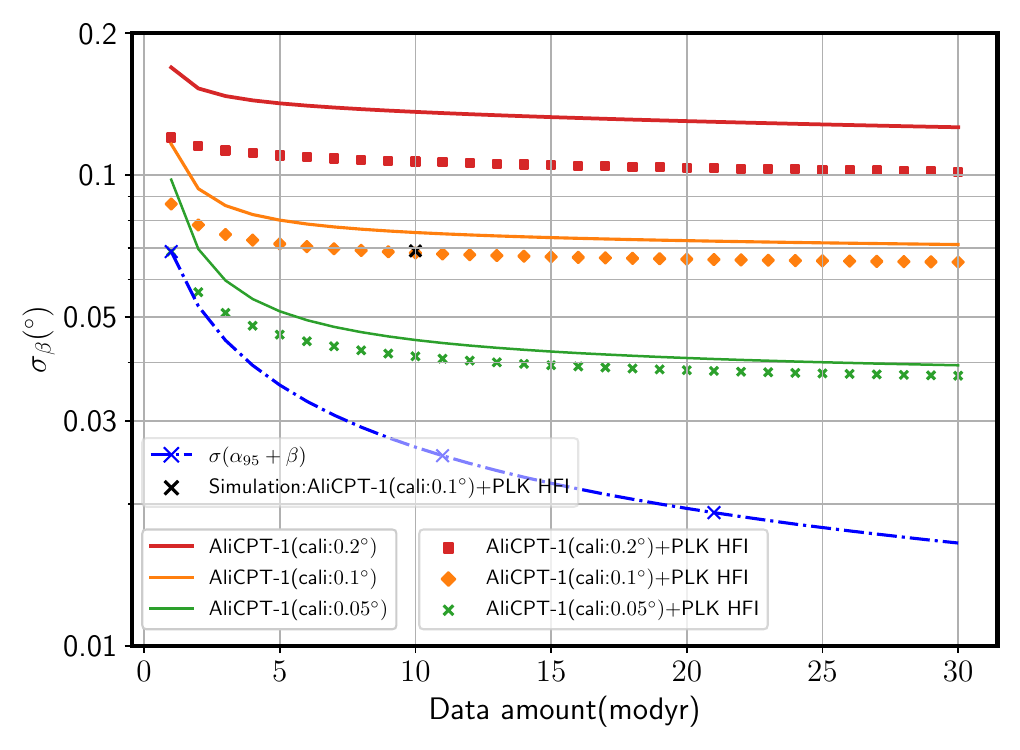}
	\caption{Evolution of $\sigma_\beta$ with AliCPT-1 data accumulation, under $\sigma^{cali}_\alpha = 0.2^\circ, 0.1^\circ, 0.05^\circ$ for AliCPT-1 bands.}\label{fig:sig_beta_vs_noise}
\end{figure}

Finally, we investigate how  $\sigma_\beta$ evolves with the AliCPT-1 noise level under different values of  $\sigma^{cali}_\alpha$, as shown in Fig.~\ref{fig:sig_beta_vs_noise}. The main findings are as follows:
\begin{itemize}
    \item Under the current calibration precision, improving $\sigma^{cali}_\alpha$ provides a more effective way for strengthen the constraint on  $\sigma_\beta$ than accumulating more CMB data.
    
    \item The curve of $\sigma(\alpha+\beta)$ approximately dominates the best possible $\sigma_\beta$, even with a perfect calibration.
    
    \item We performed simulations for the case of $\sigma^{cali}_\alpha = 0.1\si{\degree}$ and $10$ module-years of AliCPT-1 data(see  Appendix.~\ref{append:simulation}). The resulting $\sigma_\beta$ remains consistent with that from the fisher forecast .
    
    \item For our case, with input of $\beta = 0.35\si{\degree}$, a $5\sigma$ detection requires about $10$ module-years of AliCPT-1 data combined with \textit{Planck} HFI data.
\end{itemize}

\subsection{{Discussion}}
\label{sec:bias_study}

In previous sections, we studied the uncertainty of the polarization rotation angle without incorporating any foreground template. However, intrinsic foreground EB and various systematic effects—such as intensity-to-polarization leakage, beam leakage, or cross-polarization effects—can introduce significant bias into the estimation of the polarization rotation angle in the Minami-Komatsu formalism. Simulations in Appendix B show that  neglecting foreground $EB$ and  imposing no calibration prior leads to a bias of $-0.18^\circ(0.8\sigma)$ in $\beta$. We emphasize that the inclusion of calibration prior information can substantially mitigate this bias into $\sim -0.03(0.4\sigma)$. In addition, Ref.~\cite{Diego-Palazuelos:2022cnh} suggests that employing an accurate foreground template can further reduce both the bias and the uncertainty. The improvement in uncertainty arises because the correlation between the foreground template and the data gives a negative contribution to the covariance matrix (see Appendix D of Ref.~\cite{Diego-Palazuelos:2022cnh}). We will investigate the impact of foreground templates for the AliCPT-1+Planck HFI combined data which is present in Appendix~\ref{append:simulation}.

The simulation setup is similar to that described in Appendix~\ref{append:simulation}, except that a foreground template is included in this case. To investigate the effects of different templates, we consider three PySM models: a low-complexity model (‘d9’, ‘s4’, ‘f1’, ‘a1’, ‘co1’), a medium-complexity model—identical to that used in the data simulations—and a high-complexity model (‘d12’, ‘s7’, ‘f1’, ‘a2’, ‘co3’). In the likelihood analysis based on (Eq.~\ref{eq:eb_multi_likelihood_bined}), the amplitude parameter $\mathcal{A}$ is varied simultaneously with polarization rotation angle parameters.

\begin{figure}[htbp]
 \centering    
	\includegraphics[width=0.9\columnwidth]{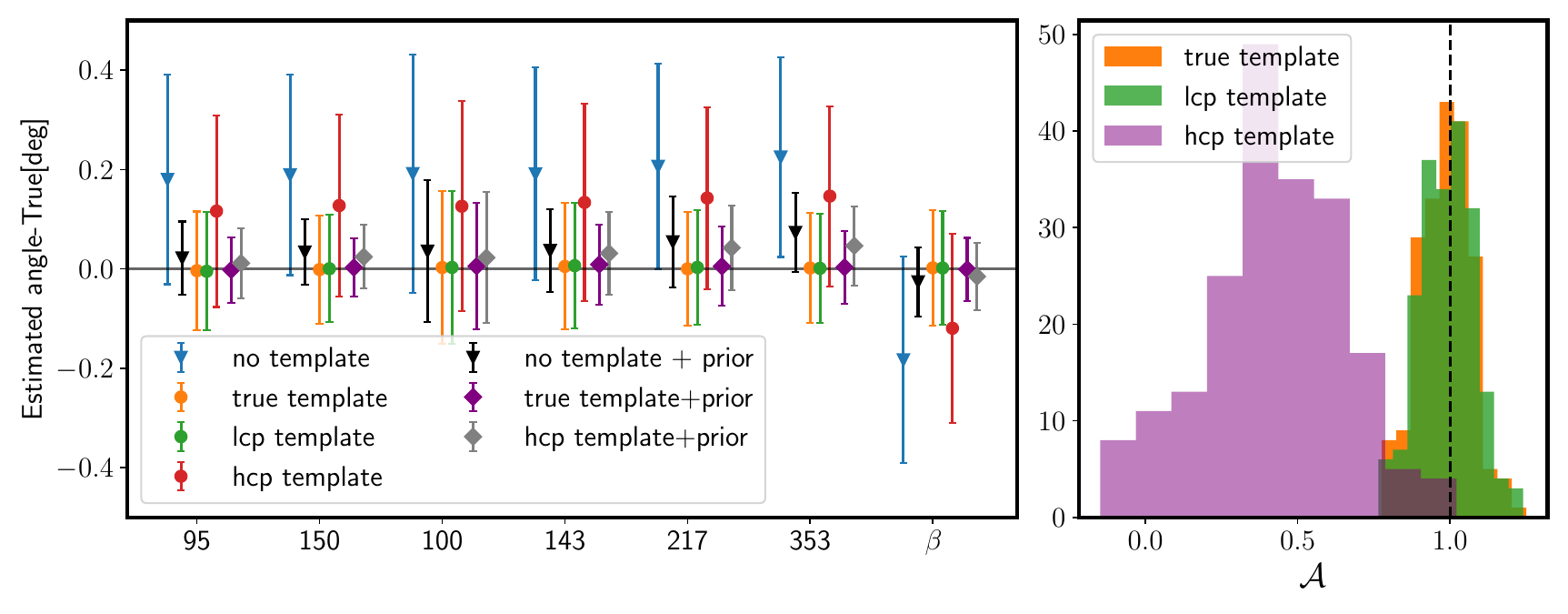}
	\caption{Left:Simulation results for the polarization rotation angle parameters derived from AliCPT-1(10 module-years of observation)+Planck HFI data, with and without foreground EB templates. "lcp" denotes low complexity foreground model, while "hcp" represents high complexity model. The error bars indicate the dispersion across simulations. "prior" indicates the application of a $0.1\si{\degree}$ prior on the polarization miscalibration angles of the two AliCPT-1 bands. Right: Distribution of $\mathcal{A}$ from 200 simulations.}\label{fig:fgEBbias}
\end{figure}

We perform a minimum likelihood estimation on 200 simulations, and the results are present in Figure.~\ref{fig:fgEBbias}. For comparison, we also show the results obtained without any foreground template(see also the Figure.~\ref{fig:triangle_const_angle} in Appendix~\ref{append:simulation}). When the template closely matches the foregrounds in the simulated data, the bias in $\beta$ becomes nearly negligible, at the level of $\sim0.002\si{\degree}(0.02\sigma)$, and the uncertainty is reduced to $0.11^\circ$ which is nearly half of value obtained without using a template. However, if the template differs significantly from the foregrounds present in the data—for example, when the high-complexity model is used as the template—both the bias ($\sim -0.12\si{\degree}, 0.6\sigma$ ) and uncertainty($\sim0.19\si{\degree}$) in $\beta$ become comparable to that with no template and no calibration prior. This highlights the importance of accurate foreground modeling. The amplitude parameter $\mathcal{A}$ is expected to be consistent with unity if the template is correct; otherwise, it will deviate from unity, as shown in the right panel of Figure.~\ref{fig:fgEBbias}.

\begin{figure}[htbp]
	\centering
		\includegraphics[width=0.6\columnwidth]{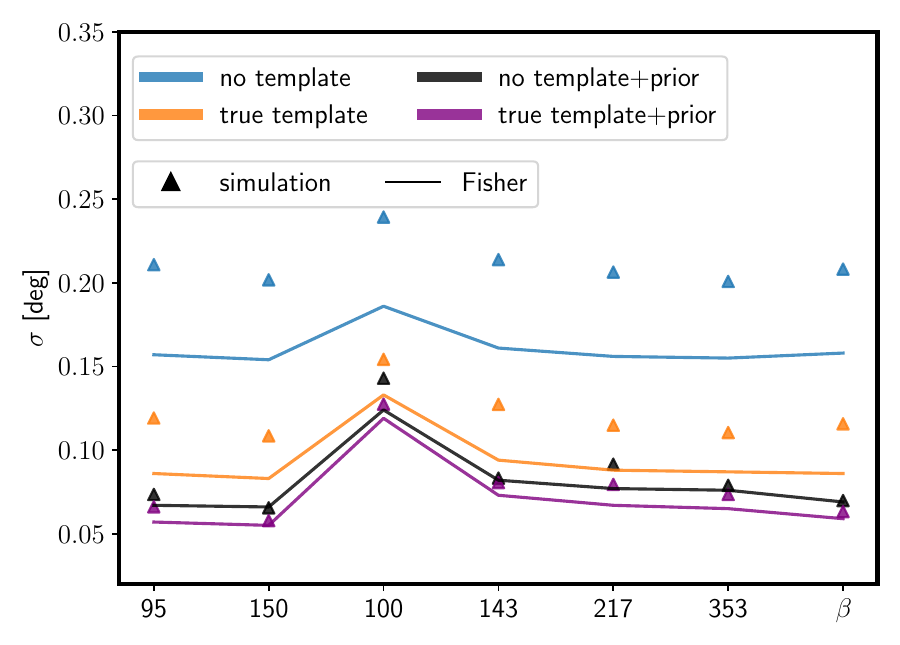}
		\caption{Comparison of the polarization rotation angle uncertainties derived from the simulation  dispersion and from the Fisher forecast. The data configuration is the same as that in Figure.~\ref{fig:fgEBbias}}
		\label{fig:fgEBsigma}
\end{figure}

Using an accurate foreground template in the analysis is more effective than applying calibration priors alone for reducing the bias in the polarization rotation angles; however, the latter performs better in reducing the uncertainty. As show in Figure.~\ref{fig:fgEBsigma}, in the absence of a prior,  the simulation-based uncertainties significantly exceed the Fisher-forecast lower bound values. Once the calibration prior is imposed, the former become very close to the latter. A joint analysis that combines both a foreground template and calibration priors is therefore the optimal choice. Even when an incorrect “hcp” foreground template is adopted in the joint analysis, the simulations yield a bias and uncertainty in $\beta$ of only $-0.015\si{\degree}(0.2\sigma)$ and $0.068\si{\degree}$ respectively, both smaller than those obtained using calibration prior alone. With an accurate foreground template, this approach delivers the best overall performance among all cases shown in Figure.~\ref{fig:fgEBbias}, achieving both a negligible bias, $\sim -0.0008\si{\degree}(0.01\sigma)$, and the smallest uncertainty,$\sim 0.063\si{\degree}$ for the AliCPT-1(10 modules-years of observation) and Planck HFI combined data.

Systematic effects generally require simulations at the time-ordered data (TOD) level. Since the analysis in this paper is based on simulated power spectra, such effects are not considered here. As a complementary check, we investigate the impact of noise inhomogeneity on the polarization rotation angle using simulations that include only CMB and noise, without polarization rotation. The noise simulation adopted here differs from that described in Appendix~\ref{append:simulation}. Specifically, rather than using the noise levels listed in Table.~\ref{tab:bands_design}, we generate the noise maps based on the pixel noise variance map shown in Figure.~\ref{fig:wide_scan}.

\begin{figure}[htbp]
	\centering    
	\includegraphics[width=0.6\columnwidth]{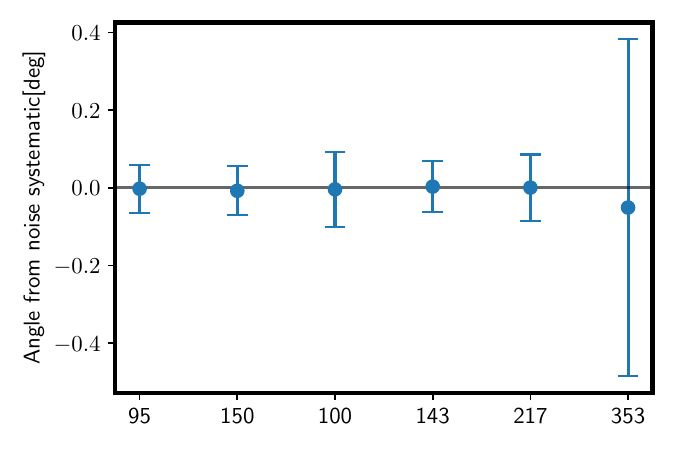}
	\caption{Polarization rotation angle estimation based in CMB + Noise simulations. The uncertainty is calculated as the dispersion across simulations. }\label{fig:noisebias}
\end{figure}

In case of CMB + noise simulation, there are no foregrounds so $\alpha$ and $\beta$ cannot be distinguished. by setting $\beta=0$ in the equations, we effectively estimate $\bar{\alpha}_{sys}=\alpha+\beta$. The results, shown in the Figure.~\ref{fig:noisebias}, indicate that the noise levels in AliCPT’s 95 and 150~GHz frequency bands do not introduce any significant bias, $\bar{\alpha}_{sys,95}=-0.002\pm0.06^\circ$ and  $\bar{\alpha}_{sys,150}=-0.008\pm0.06^\circ$. The uncertainty in $\bar{\alpha}_{sys}$ for each frequency band is consistent with the corresponding noise level. A more detailed investigation on the systematic effects is left for future work.

\section{Forecast for Anisotropic Polarization Rotation}\label{sec:forecast_aniso}

In this section, we present a forecast for AliCPT's sensitivity to anisotropic polarization rotation.
Following the formalism and notation of Ref.~\cite{Zhong:2024tgw}, we reconstruct the rotation field using the quadratic estimator technique applied to mock CMB polarization data. 
Previous studies have shown that large-aperture telescopes provide significant advantages in measuring anisotropic polarization rotation, owing to their higher angular resolution and improved sensitivity to small-scale polarization structures. 
Motivated by this, in addition to the AliCPT-1 configuration described in the previous section, we also consider a possible large-aperture telescope, referred to as AliCPT-LAT, featuring a $6\,\mathrm{m}$ aperture and adopting the same detector module design as AliCPT-1. 

\subsection{Methodology}

The anisotropic polarization rotation can be described as a direction-dependent rotation angle $\beta(\hat{\boldsymbol{n}})$ acting on the Stokes parameters of the CMB. Expanding to first order in $\beta$, the induced perturbations on the E- and B-mode coefficients are given by
\begin{equation}
  \label{eq:1st E and B rotation}
  \begin{aligned}
  \delta E'_{\ell m} &= -2\sum_{LM}\sum_{\ell'm'} (-1)^{m}\beta_{LM} 
  \begin{pmatrix}
  \ell & L & \ell' \\
  -m & M & m'
  \end{pmatrix} 
  {}_{2} F^{\beta}_{\ell L \ell'} \eta_{\ell L \ell'} E_{\ell' m'},\\
  \delta B'_{\ell m} &=\ 2\sum_{LM}\sum_{\ell'm'} (-1)^{m}\beta_{LM} 
  \begin{pmatrix}
  \ell & L & \ell' \\
  -m & M & m'
  \end{pmatrix}
  {}_{2} F^{\beta}_{\ell L \ell'} \epsilon_{\ell L \ell'} E_{\ell' m'},
  \end{aligned}
\end{equation}
where the spatial fluctuation of the rotation angle $\beta(\hat{\boldsymbol{n}})$ is expanded in spherical harmonics as
\begin{equation}
    \beta(\hat{\boldsymbol{n}})=\sum_{LM} \beta_{LM} Y_{LM}(\hat{\boldsymbol{n}}),
\end{equation}
and the parity-dependent factors are defined by
\begin{equation}
  \label{eq:parity}
  \begin{aligned}
  &\eta_{\ell L \ell'}\equiv\frac{1-(-1)^{\ell+L+\ell'}}{2 i}, \\
  &\epsilon_{\ell L \ell'}\equiv\frac{1+(-1)^{\ell+L+\ell'}}{2},
  \end{aligned}
\end{equation}
while the geometrical coupling coefficient reads
\begin{equation}
{}_{2} F^{\beta}_{\ell L \ell'} = 
\sqrt{\frac{\left(2 \ell+1\right)(2L+1)\left(2 \ell'+1\right)}{4\pi}}
\begin{pmatrix}
\ell & L & \ell' \\
2 & 0 & -2
\end{pmatrix}.
\end{equation}

Eqs.~(\ref{eq:1st E and B rotation}) describe how $\beta(\hat{\boldsymbol{n}})$ mixes the primary E-modes into B-modes, producing characteristic off-diagonal correlations between different multipoles. The unnormalized quadratic estimator is defined as
\begin{equation}
   \label{eq:unormalized estimator}
\bar{\beta}_{LM}= 
\sum_{\ell _{1} m_{1}} 
\sum_{\ell _{2}m_{2}} 
(-1)^{M} 
\begin{pmatrix} 
\ell_1  & \ell _{2}  & L \\ 
m_{1}  &  m_{2}  &  -M
\end{pmatrix} 
f_{\ell_{1} \ell_{2} L}^\beta
\frac{E_{\ell_{1 }m_{1}}}{C^{EE}_{\ell_{1}}} 
\frac{B_{\ell_{2} m_{2}}}{C^{BB}_{\ell_{2}}},
\end{equation}
and the normalized, unbiased estimator is obtained as
\begin{equation}
    \label{eq:normailized estimator}
   \hat{\beta}_{LM}=A_{L}\left(\bar{\beta}_{LM}-\langle\bar{\beta}_{LM}\rangle\right),
\end{equation}
where weighting functions $f^{\beta}_{\ell L \ell'} = -2\epsilon_{\ell L \ell'}{}_{2} F^{\beta}_{\ell L \ell'}C^{\mathrm{EE}}_{\ell}$, and
$A_{L}$ is the normalization factor ensuring an unbiased reconstruction. A detailed discussion can be found in Ref.~\cite{Zhong:2024tgw}.

The angular power spectrum of the reconstructed rotation field is then estimated by
\begin{equation}
\label{eq:ps estimator}
    \hat{C}^{\beta\beta}_L= C^{\hat{\beta} \hat{\beta}}_L 
    - {}^{(\mathrm{RD})}N^{(0)}_L 
    - N^{(1)}_L 
    - N^{\mathrm{Lens}}_L.
\end{equation}
Here, the first term $C^{\hat{\beta}\hat{\beta}}_L$ represents the raw spectrum of the reconstructed field, while ${}^{(\mathrm{RD})}N^{(0)}_L$ is the disconnected Gaussian noise bias estimated from random simulations, $N^{(1)}_L$ accounts for the secondary contraction bias, and $N^{\mathrm{Lens}}_L$ denotes the contamination from the CMB lensing potential. We combine the estimators of AliCPT two bands in harmonic space. All of these quantities are all computed within the same simulation framework to ensure unbiased power-spectrum estimation.

The scale-invariant power spectrum of anisotropic rotation is defined as:
\begin{equation}
    \frac{L(L+1)}{2\pi}C_L^{\beta\beta}\equiv A_{\mathrm{CB}}\times10^{-4}\text{ [rad}^2],
\end{equation}
where the $A_{\mathrm{CB}}$ is estimated by HL log likelihood\cite{Hamimeche:2008ai}.

\subsection{Simulation and Result}

We follow the same simulation procedure as detailed in Ref.~\cite{Zhong:2024tgw} to generate mock data for the CMB, the rotation angle field, and the lensing potential. 
The simulated sky maps are produced under identical scanning strategies and noise realizations, and the same mask is applied to remove regions near the Galactic plane and the edges of the survey. 
After masking, the effective sky fraction reaches $f_{\mathrm{sky}} = 23.6\%$.

To mitigate the contamination from Galactic foregrounds, the multipoles below $\ell \leq 200$ in the simulated maps are removed in advance. 
In the quadratic estimator, only the range $20 < L < 1200$ is retained for the reconstruction of the anisotropic polarization rotation angle, as the largest-scale modes ($L \lesssim 20$) are affected by the isotropic polarization rotation\cite{Namikawa:2020ffr}.

\begin{figure}[htbp]
	\centering    
    \includegraphics[width=0.8\columnwidth]{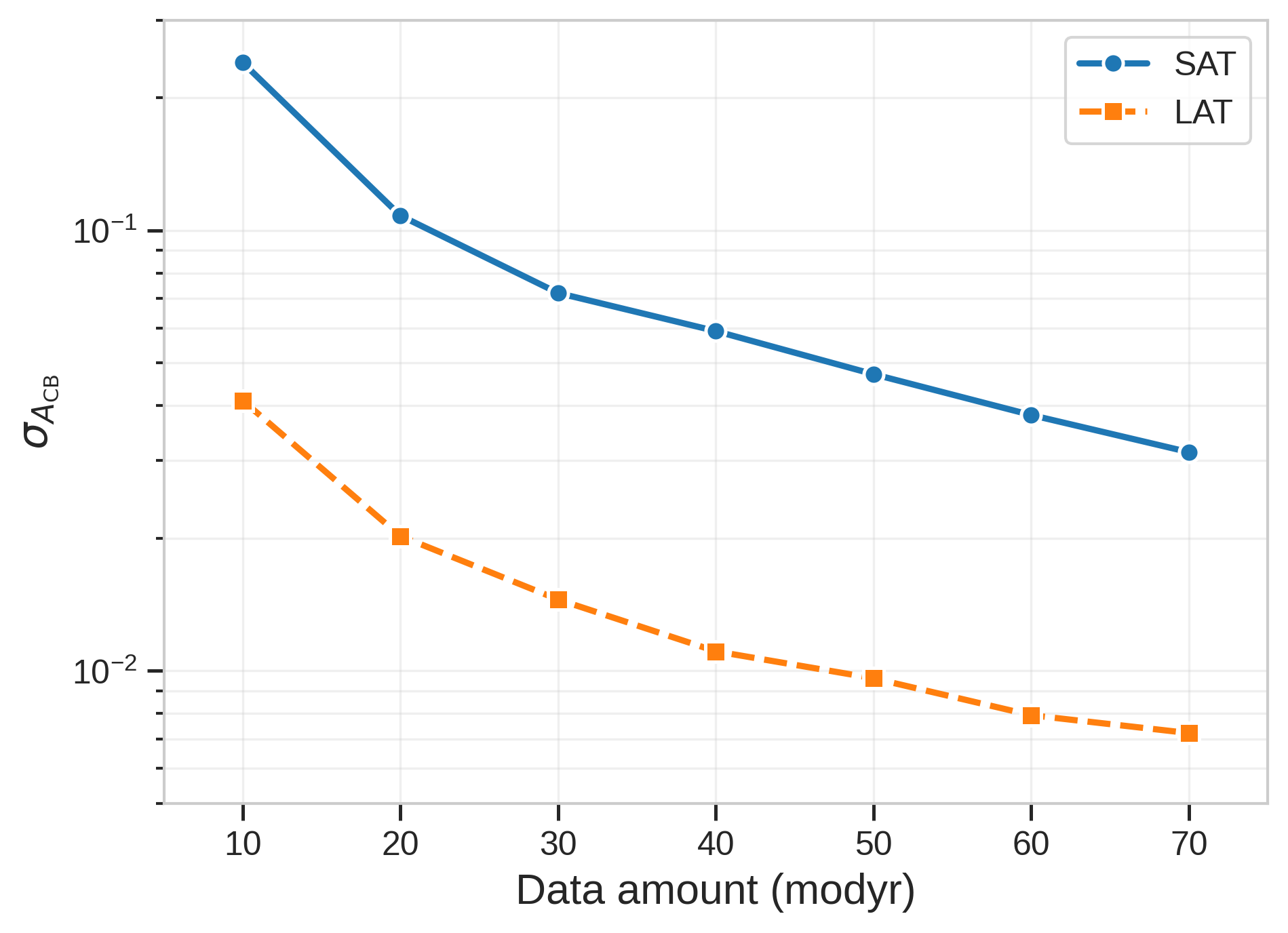}
	\caption{Forecasted sensitivity of the anisotropic polarization rotation amplitude $A_{\mathrm{CB}}$ as a function of AliCPT data accumulation (module-years).}
	\label{fig:constraint_a_cb}
\end{figure}

Figure.~\ref{fig:constraint_a_cb} shows the forecasted sensitivity to $A_{\mathrm{CB}}$, as a function of the cumulative observing time (in module-year). The results are presented for both the baseline AliCPT-1, which is a small-aperture telescope (SAT) and the possible large-aperture telescope (LAT) configurations.

As shown in the figure, the sensitivity approximately follows a power-law trend. 
The LAT exhibits a substantial enhancement in performance, achieving about a factor of six better sensitivity than the SAT for the same data amount. 
For instance, with a total data amount of $50$~module-year, the expected $1\sigma$ uncertainty for the SAT configuration reaches $\sigma_{A_{\mathrm{CB}}}\sim 4.7 \times10^{-2}$, corresponding roughly to the current $2\sigma$ upper limit of $A_{\mathrm{CB}}\simeq4.4\times10^{-2}$ obtained by BICEP/Keck.
In contrast, the LAT achieves $\sigma_{A_{\mathrm{CB}}} \sim 9 \times 10^{-3}$, demonstrating the substantial advantage of a large-aperture system in probing spatially dependent polarization rotation under realistic noise and sky-coverage assumptions.
This improvement originates from the higher angular resolution, which allows for more effective reconstruction of small-scale features in the polarization rotation field.

\section{Summary}\label{sec:summary}

Recent results from the DESI survey provide compelling evidence for the dynamical nature of dark energy, with the equation of state parameter showing a transition across $w=-1$. 
Motivated by these findings, we investigate the interaction between dark energy and photons through the Chern-Simons coupling, which induces a rotation of the CMB polarization plane. 
By measuring the $TB$ and $EB$ power spectra of the CMB, this effect offers a microphysical probe of dynamical dark energy, complementing traditional approaches based on its gravitational influence.

In this work, we develop a Fisher-matrix framework to forecast the joint estimation of the instrumental polarization miscalibration angle $\alpha$ and the CMB isotropic rotation angle $\beta$. 
Our analysis incorporates the use of external calibration together with foreground radiation to break the degeneracy between $\alpha$ and $\beta$. 
We find that including external calibration substantially improves the constraint on $\beta$. 
{Assuming a fiducial value of $\beta = 0.35\si{\degree}$ and a calibration accuracy of $0.1^\circ$,  the combination of 10 module-years of AliCPT observations with \textit{Planck} HFI data yields $\hat{\beta}= 0.32\si{\degree} \pm 0.069\si{\degree}$ in our simulations, corresponding to a nominal detection significance close to $5\sigma$. If a sufficiently accurate foreground template is available, the result further improves to  $\hat{\beta} = 0.35\si{\degree}\pm 0.063\si{\degree}$, corresponding to a significance of about $\sim 5.5\sigma$.} This would open a new window for probing the interaction between dark energy and ordinary matter, as well as for investigating the dynamical nature of dark energy~\cite{Lin:2025gne}. 

Furthermore, we investigate AliCPT's potential for detecting anisotropies in the polarization rotation angle. Unlike constraints on the isotropic rotation angle, a large-aperture telescope can improve constraints on the anisotropic rotation angle by a factor $\sim 6$. With a large-aperture configuration and 50 module-years of observation, AliCPT is projected to reach the sensitivity twice better than the current best limit. 

\section*{Acknowledgments}
We acknowledge
the use of CAMB package in calculating the power spectra, Healpy package for map simulation, PYSM package for foreground simulation and NaMaster for power spectrum estimation. We thank Sebastian Belkner for discussion.
This work is supported by the National Natural Science Foundation of China
No.12403005, {CAS Project for Young Scientists in Basic Research No.YSBR-138} and the National Key R\&D Program of China No. 2020YFC2201601.

\appendix

\section{{Discussion on covariance matrix}}
\label{append:covariancce}

The calculation of covariance matrix $\Xi$ in Eq.~(\ref{eq:cov_exp}) depends on the $Q$ covariance matrix of power spectrum vectors $\vec{C}_\ell^{ij, o/c/f}$. Except for $Q^{iji'j', o*o}$, the explicit expressions for these matrices are provided below,
\begin{eqnarray}\label{eq:qother}
Q^{iji'j',f*f} &=& \left( {\begin{array}{cc}
	\mathrm{Cov}\Big( C_\ell^{E^{i}B^{j}, fg}, C_\ell^{E^{i'}B^{j'}, fg}\Big)	& \mathrm{Cov}\Big( C_\ell^{E^{i}B^{j}, fg}, C_\ell^{B^{i'}E^{j'}, fg}\Big)  \\
	\mathrm{Cov}\Big( C_\ell^{B^{i}E^{j}, fg}, C_\ell^{E^{i'}B^{j'}, fg}\Big)	& \mathrm{Cov}\Big( C_\ell^{B^{i}E^{j}, fg}, C_\ell^{B^{i'}E^{j'}, fg}\Big)\\
	\end{array} } \right),\nn\\	
Q^{iji'j',o*f} &=& \left( {\begin{array}{cc}
	\mathrm{Cov}\Big( C_\ell^{E^{i}E^{j}, o}, C_\ell^{E^{i'}B^{j'}, fg}\Big)	& \mathrm{Cov}\Big( C_\ell^{E^{i}E^{j}, o}, C_\ell^{B^{i'}E^{j'}, fg}\Big)  \\
	\mathrm{Cov}\Big( C_\ell^{B^{i}B^{j}, o}, C_\ell^{E^{i'}B^{j'}, fg}\Big)	& \mathrm{Cov}\Big( C_\ell^{B^{i}B^{j}, o}, C_\ell^{B^{i'}E^{j'}, fg}\Big)\\
	\mathrm{Cov}\Big( C_\ell^{E^{i}B^{j}, o}, C_\ell^{E^{i'}B^{j'}, fg}\Big)	& \mathrm{Cov}\Big( C_\ell^{E^{i}B^{j}, o}, C_\ell^{B^{i'}E^{j'}, fg}\Big)\\		
	\end{array} } \right),\nn\\
Q^{iji'j',f*o} &=& \left( {\begin{array}{ccc}
	\mathrm{Cov}\Big( C_\ell^{E^{i}B^{j}, fg}, C_\ell^{E^{i'}E^{j'}, o}\Big)	& \mathrm{Cov}\Big( C_\ell^{E^{i}B^{j}, fg}, C_\ell^{B^{i'}B^{j'}, o}\Big) &  \mathrm{Cov}\Big( C_\ell^{E^{i}B^{j}, fg}, C_\ell^{E^{i'}B^{j'}, o}\Big) \\
	\mathrm{Cov}\Big( C_\ell^{B^{i}E^{j}, fg}, C_\ell^{E^{i'}E^{j'}, o}\Big)	& \mathrm{Cov}\Big( C_\ell^{B^{i}E^{j}, fg}, C_\ell^{B^{i'}B^{j'}, o}\Big) &  \mathrm{Cov}\Big( C_\ell^{B^{i}E^{j}, fg}, C_\ell^{E^{i'}B^{j'}, o}\Big)\\
	\end{array} } \right), \nn\\
\end{eqnarray}
the auto correlated $Q$ matrices have symmetry $Q^{pq, T} = Q^{qp}$, while the cross correlated $Q$ matrices share conjugate symmetry $Q^{pq, f*o, T} = Q^{qp, o*f}$.

In fisher forecast, we give the analytical expression of the $Q$ covariance matrix between the  foreground and the observed data.  If we assume no correction between the foreground and the CMB, $Q^{o*f}$ reduces to $Q^{rf*f}$ where $rf$ denotes the rotated foreground. The rotated foreground power spectrum are obtained by,   
\bea\label{rotated_spectra_ij_sim}
{C}_l^{E^iB^j, r} &=& C_\ell^{E^iE^j}\cos(2\alpha^i)\sin(2\alpha^j)-C_\ell^{B^iB^j}\sin(2\alpha^i)\cos(2\alpha^j) \nn\\
&& + C_\ell^{E^iB^j}\cos(2\alpha^{i})\cos(2\alpha^{j}) -C_\ell^{E^jB^i}\sin(2\alpha^{i})\sin(2\alpha^{j}) , \nn\\
{C}_l^{E^iE^j, r} &=& C_\ell^{E^iE^j}\cos(2\alpha^i)\cos(2\alpha^j)+C_\ell^{B^iB^j}\sin(2\alpha^i)\sin(2\alpha^j) \nn\\
&& - C_\ell^{E^iB^j}\cos(2\alpha^{i})\sin(2\alpha^{j})-C_\ell^{E^jB^i}\cos(2\alpha^{j})\sin(2\alpha^{i}), \nn\\
{C}_l^{B^iB^j,r} &=& C_\ell^{E^iE^j}\sin(2\alpha^i)\sin(2\alpha^j)+C_\ell^{B^iB^j}\cos(2\alpha^i)\cos(2\alpha^j)\nn\\
&& +C_\ell^{E^iB^j}\sin(2\alpha^{i})\cos(2\alpha^{j})+C_\ell^{E^jB^i}\sin(2\alpha^{j})\cos(2\alpha^{i}), 
\eea
Rewrite in matrix form, we have
\bea
&& \left( {\begin{array}{c}
		C_\ell^{E^{i}E^{j}, rfg} \\
		C_\ell^{B^{i}B^{j}, rfg}	\\
		C_\ell^{E^{i}B^{j}, rfg}\\		
\end{array} } \right)= \mathcal{R}_f^{ij}\vec{\mathbb{C}}^{ij,f }_\ell, ~~ \vec{\mathbb{C}}^{ij,f }_\ell=  \left( {\begin{array}{c}
		C_\ell^{E^{i}E^{j}, fg} \\
		C_\ell^{B^{i}B^{j}, fg}	\\
		C_\ell^{E^{i}B^{j}, fg}\\	
		C_\ell^{B^{i}E^{j}, fg}\\	
\end{array} } \right), \nn\\
&& \mathcal{R}_f^{ij} = \left( {\begin{array}{cccc}
		\cos(2\alpha^i)\cos(2\alpha^j)  & \sin(2\alpha^i)\sin(2\alpha^j) & -\cos(2\alpha^i)\sin(2\alpha^j) & -\sin(2\alpha^i)\cos(2\alpha^j) \\
		\sin(2\alpha^i)\sin(2\alpha^j) &\cos(2\alpha^i)\cos(2\alpha^j) &\sin(2\alpha^i)\cos(2\alpha^j) &\cos(2\alpha^i)\sin(2\alpha^j)	\\
		\cos(2\alpha^i)\sin(2\alpha^j) & -\sin(2\alpha^i)\cos(2\alpha^j) & \cos(2\alpha^i)\cos(2\alpha^j) &-\sin(2\alpha^i)\sin(2\alpha^j)  \\		
\end{array} } \right)\nn\\
\eea

Then for the cross correlation part between the foreground and the observed data in the covariance matrix $\Xi$, we have
\bea\label{eq:cov_rotfg_fg}
&& \Xi^{pq, o*fg}_\ell = \mathrm{Cov}\left(\vec{A}_o^{ij, T} \vec{C}^{ij, rfg}_\ell,  \vec{A}_f^{i'j', T} \vec{C}^{i'j', fg}_\ell \right) 
= \vec{A}_o^{ij, T} \mathcal{R}_f^{ij}  \mathrm{Cov}\left(\vec{{C}}^{ij,f}_\ell,  \vec{C}^{i'j', fg, T}_\ell\right)  \vec{A}_f^{i'j'}     \nn\\ 
&&\mathrm{Cov}\left(\vec{{C}}^{ij,f}_\ell,  \vec{C}^{i'j', fg, T}_\ell\right) =  \left( {\begin{array}{cc}
		\mathrm{Cov}\Big( {C}_\ell^{E^{i}E^{j}, fg}, C_\ell^{E^{i'}B^{j'}, fg}\Big)	& \mathrm{Cov}\Big( {C}_\ell^{E^{i}E^{j}, fg}, C_\ell^{B^{i'}E^{j'}, fg}\Big) \\
		\mathrm{Cov}\Big( {C}_\ell^{B^{i}B^{j}, fg}, C_\ell^{E^{i'}B^{j'}, fg}\Big)	& \mathrm{Cov}\Big( {C}_\ell^{B^{i}B^{j}, fg}, C_\ell^{B^{i'}E^{j'}, fg}\Big)	\\		
		\mathrm{Cov}\Big( {C}_\ell^{E^{i}B^{j}, fg}, C_\ell^{E^{i'}B^{j'}, fg}\Big)	& \mathrm{Cov}\Big( {C}_\ell^{E^{i}B^{j}, fg}, C_\ell^{B^{i'}E^{j'}, fg}\Big)\\  		
		\mathrm{Cov}\Big( {C}_\ell^{B^{i}E^{j}, fg}, C_\ell^{E^{i'}B^{j'}, fg}\Big)	& \mathrm{Cov}\Big( {C}_\ell^{B^{i}E^{j}, fg}, C_\ell^{B^{i'}E^{j'}, fg}\Big)
\end{array} } \right),\nn\\
\eea

Then we explain in Sec.~\ref{sec:Methodology} why the contribution of $\vec{T}^c_\ell$ to the covariance can be safely neglected. Since the CMB has no intrinsic $EB$ correlation, its power spectrum covariance can be simplified accordingly,
\bea
Q^{iji'j',c*c} &=& \left( {\begin{array}{cc}
		\mathrm{Cov}\Big( C_\ell^{E^{i}E^{j}, cmb}, C_\ell^{E^{i'}E^{j'}, cmb}\Big)	& \mathrm{Cov}\Big( C_\ell^{E^{i}E^{j}, cmb}, C_\ell^{B^{i'}B^{j'}, cmb}\Big)  \\
		\mathrm{Cov}\Big( C_\ell^{B^{i}B^{j}, cmb}, C_\ell^{E^{i'}E^{j'}, cmb}\Big)	& \mathrm{Cov}\Big( C_\ell^{B^{i}B^{j}, cmb}, C_\ell^{B^{i'}B^{j'}, cmb}\Big)\\
\end{array} } \right)\nn\\
&=& \frac{b^i_\ell b^{i'}_\ell b^j_\ell b^{j'}_\ell}{(2\ell+1)f_{sky}}\left( {\begin{array}{cc}
		\big( C_\ell^{EE, cmb, theo}\big)^2 & 0\\
		0	& \big( C_\ell^{BB, cmb, theo}\big)^2 \\
\end{array} } \right)\nn\\
\eea
where $b^i_\ell$ is the beam smooth function and $C_\ell^{EE/BB, theo}$ are theoretical CMB power spectrum. 

By following the same procedure of Eqs.~(\ref{rotated_spectra_ij_sim}--\ref{eq:cov_rotfg_fg}) but with $\alpha$ replaced by $\alpha+\beta$, and applying appropriate simplifications, we obtain,
\bea
&&\Xi^{pq, o*c}_\ell = \Xi^{pq, c*o}_\ell  = -\Xi^{pq, c*c}_\ell \nn\\
& =&  -\frac{\sin^2(4\beta)b^i_\ell b^{i'}_\ell b^j_\ell b^{j'}_\ell}{2\cos(2\alpha^i+2\alpha^j)\cos(2\alpha^{i'}+2\alpha^{j'})}\left[ \big( C_\ell^{EE, cmb, theo}\big)^2 + \big( C_\ell^{BB, cmb, theo}\big)^2\right]
\eea

\begin{figure}[htbp]
	\centering
	\includegraphics[width=0.9\columnwidth]{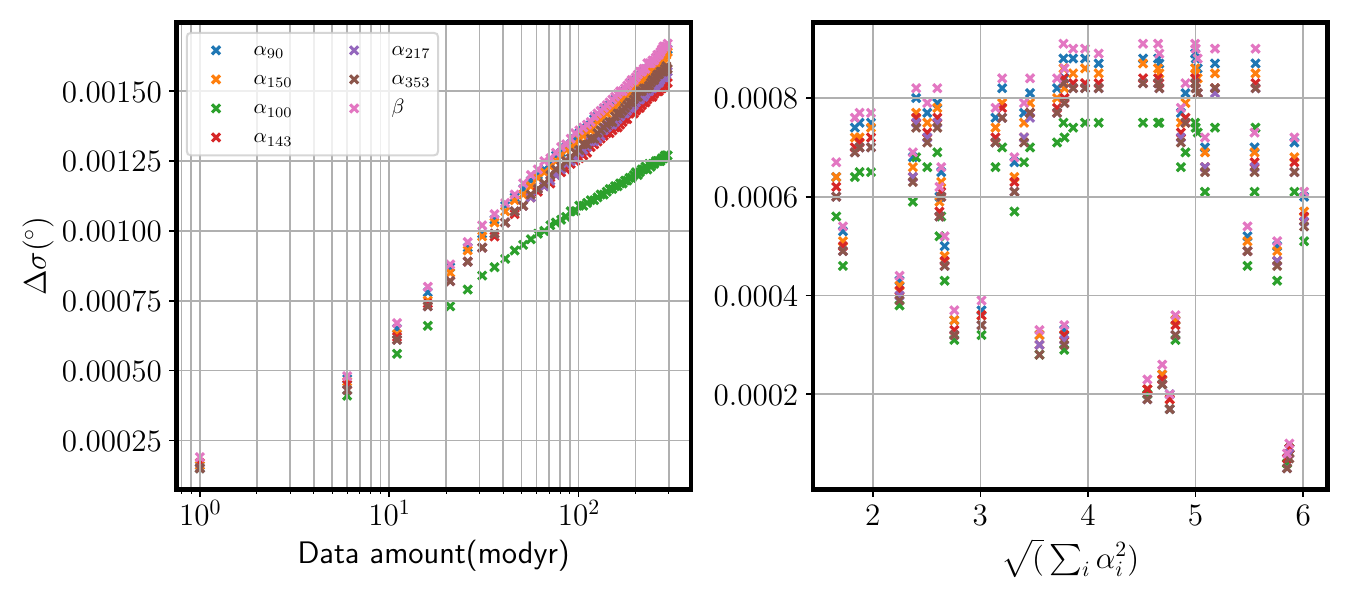}
	\caption{The difference in Fisher-forecasted uncertainties of the polarization angle—with and without the CMB contribution to the covariance—is evaluated for the AliCPT-1+\textit{Planck} HFI dataset, without a calibration prior and the foreground template. Left: Results for different noise levels(data amount); Right: Results for different $\alpha$ values, sampled from the random polarization angles present in Figure.~\ref{fig:input_rotation}.}\label{fig:cov_cmb_test}
\end{figure}

The numerical test result is present in Figure.~\ref{fig:cov_cmb_test}. As shown in the left panel, the difference in polarization angle uncertainty—with and without including the CMB contribution in the covariance matrix—increases as the data amount accumulates logarithmically. This is expected, since the CMB contribution becomes increasingly important at lower noise levels. However, for noise levels corresponding to tens of module-years of observation for AliCPT-1, the resulting increase in polarization angle uncertainty is less than $10^{-3}$ deg, which is negligible for the purposes of this work. The same conclusion applies when varying the input polarization rotation angle values, as illustrated in the right panel.

\section{{Derivation of the Fisher matrix}}
\label{append:fisher}

The evaluation of Eqs.~(\ref{eq:fisher_matrix_expression}) depends on the values of derivatives of both covariance matrix $\Xi$ and the coefficient vectors $\vec{A}_{o/c/f}$ with respect to the polarization rotation angle parameters $\alpha_i$ and $\beta$.
For the covariance matrix, from its definition expression in Eqs.~(\ref{eq:cov_exp}, \ref{eq:cov_def}), the partial derivatives are obtained via the Leibnitz rule as
\begin{eqnarray}\label{eq:dl_matrix_def}
\Xi_\ell^{pq}&=& \Xi_\ell^{pq, o*o} +  \Xi_\ell^{pq, f*f}+ \Xi_\ell^{pq, o*f} + \Xi_\ell^{pq, f*o} = \sum_{x, z \in [o, f]}  \vec{A}^{p, T}_x Q^{pq, x*z}\vec{A}^{q}_z \nn\\
&=&  \vec{A}^{p, T}_o Q^{pq, o*o}\vec{A}^{q}_o    + \vec{A}^{p, T}_f Q^{pq, f*f}\vec{A}^{q}_f + \vec{A}^{p, T}_o Q^{pq, o*f}\vec{A}^{q}_f+ \vec{A}^{p, T}_f Q^{pq, f*o}\vec{A}^{q}_o,   \nn\\
\Xi_{pq, \theta}  &=&  D^L_{pq, \theta} + D^R_{pq, \theta},~~ \theta \in [\alpha^t,\beta,\mathcal{A}]\nn\\
D^L_{pq, \alpha^t} &=&\sum_{x \in [o, f]}\left(\vec{A}^{p, T}_{x, \alpha^t} Q^{pq, x*x}\vec{A}^{q}_x \right) + \vec{A}^{p, T}_{o, \alpha^t} Q^{pq, o*f}\vec{A}^{q}_f +\vec{A}^{p, T}_{f, \alpha^t} Q^{pq, f*o}\vec{A}^{q}_{o}, \nn\\
D^R_{pq, \alpha^t} &=& \sum_{x \in [o, f]} \left\{\vec{A}^{p, T}_{x} Q^{pq, x*x}\vec{A}^{q}_{x, \alpha^t} \right\}  + \vec{A}^{p, T}_{o} Q^{pq, o*fg}\vec{A}^{q}_{f, \alpha^t} +\vec{A}^{p, T}_f Q^{pq, fg*o}\vec{A}^{q}_{o, \alpha^t}, \nn\\
D^L_{pq, \mathcal{A}} &=&\vec{A}^{p, T}_{f, \mathcal{A} } Q^{pq, f*f}\vec{A}^{q}_f +  \vec{A}^{p, T}_{f, \mathcal{A}} Q^{pq, f*o}\vec{A}^{q}_o,\nn\\
D^R_{pq, \mathcal{A}} &=&\vec{A}^{p, T}_{f } Q^{pq, f*f}\vec{A}^{q}_{f, \mathcal{A}} +  \vec{A}^{p, T}_{o} Q^{pq, o*f}\vec{A}^{q}_{f, \mathcal{A}}
\end{eqnarray}
where the derivative of covariance matrices are separated into two parts according to the Leibnitz rule and these two parts satisfy$
D^L_{pq, \theta} = D^R_{qp, \theta}$
for all $\theta$ parameters. $\vec{A}^{p}$ and $Q^{pq}$  respectively denote the coefficient vector and the covariance matrix of the observed power spectra, as defined in Section~\ref{sec:Methodology}. $p, q$ are elements indices.
The derivatives of the coefficient vectors $\vec{A}^{p}_{o/c/f}$ with respect to all parameters are given by
\bea
\vec{A}^{p}_{o, \alpha^t} &=& \frac{1}{\left[\cos(4\alpha^i)+\cos(4\alpha^j)  \right]^2}\Big\{ -4\left(1+\cos(4\alpha^i)\cos(4\alpha^j)\right)\delta_{jt} - 4\sin(4\alpha^i)\sin(4\alpha^j)\delta_{it},\nn\\ &&4\left(1+\cos(4\alpha^i)\cos(4\alpha^j)\right)\delta_{it} + 4\sin(4\alpha^i)\sin(4\alpha^j)\delta_{jt} , 0\Big\}^T, \nn\\
\vec{A}^{p}_{c, \alpha^t} &=&  \frac{\sin(4\beta)\tan2(\alpha^{i}+\alpha^{j})}{\cos2(\alpha^{i}+\alpha^{j})}(\delta_{p_it}+\delta_{p_jt})\left\{1, -1\right\}^T
, ~~\vec{A}^{p}_{c, \beta} = - \frac{2\cos(4\beta)}{\cos2(\alpha^{i}+\alpha^{j})}\left\{1, -1\right\}^T,\nn\\
\vec{A}^{p}_{f, \alpha^t} &=&   \mathcal{A}\Big\{\frac{\sin(2\alpha^i+2\alpha^j)(\delta_{p_it}+\delta_{p_jt})}{\cos^2(2\alpha^i+2\alpha^j)}  +\frac{\sin(2\alpha^i-2\alpha^j)(\delta_{p_it}-\delta_{p_jt})}{\cos^2(2\alpha^i-2\alpha^j)} ,\nn\\
&&\frac{\sin(2\alpha^i+2\alpha^j)(\delta_{p_it}+\delta_{p_jt})}{\cos^2(2\alpha^i+2\alpha^j)}  -\frac{\sin(2\alpha^i-2\alpha^j)(\delta_{p_it}-\delta_{p_jt})}{\cos^2(2\alpha^i-2\alpha^j)}\Big\}^T,\nn\\
\vec{A}^{p}_{f, \mathcal{A}} &=& \vec{A}^{p}_f(\mathcal{A}=1)
\eea
where $i, j$ corresponds to the frequency pair indices of $p-th$ element.

In the following section, we present the derivation of Eq.~(\ref{eq:fisher_matrix_expression}). For clarity, we decompose the likelihood function into three components as outlined in Eq.~(\ref{eq:eb_multi_likelihood_bined_prior}),
\beq
\mathcal{L} = \sum_{b} \left( \mathcal{L}^0_{b} +  \mathcal{L}^1_{b} \right) + \mathcal{L}^2,
\eeq
with $\mathcal{L}^1_{b}$ the logarithm of determinant of covariance matrix, and $\mathcal{L}^2_{b}$ containing the prior information.

First we derive the first order derivatives of the likelihood function with respect to the rotation angle parameter. For the three components we obtain,
\bea\label{eq:lhd_derv_0}
\mathcal{L}^0_{b,\theta} &=& \sum_{pq} \left[ \sum_x( \vec{A}^{p, T}_{x,\theta}\vec{C}^{p, x}_b )\Xi^{-1}_{b:pq} +\frac{1}{2}  \sum_x( \vec{A}^{p, T}_{x}\vec{C}^{p, x}_b ) \Xi^{-1}_{b:pq, \theta}\right] \sum_y( \vec{A}^{q, T}_{y}\vec{C}^{q, y}_b ),\nn\\
%
\mathcal{L}^1_{b,\theta} &=& \frac{1}{2} \mathrm{Tr}( \Xi^{-1}_b\Xi_{b,\theta}), ~~\mathcal{L}^2_{,\theta} = \frac{\theta- \bar{\theta}}{(\sigma_\theta^{cali})^2},
\eea
where the indices $p,q$ are summation indices. For $\mathcal{L}^0_{b}$ and $\mathcal{L}^1_{b}$, the variable $\theta$ can be any of $ \{\beta, \alpha, \mathcal{A}\}$. For $\mathcal{L}^2_b$, $\theta $ is restricted to $\alpha$. In computation of $\mathcal{L}^1_{b,\theta}$, the following formula is used,
\beq
\ln |\Xi_b| = \mathrm{Tr}(\ln \Xi_b), ~~~(\ln \Xi_b)_{, \theta} = \Xi_b^{-1}\Xi_{b, \theta}
\eeq

Before proceeding to deriving the fisher matrix expression, we examine the expectation of $\mathcal{L}_{b,\theta}$. In $\mathcal{L}^0_{b,\theta}$, the CMB power spectrum  
$\vec{C}^{o}_b$
is treated as Gaussian random variable, and similarly $\theta$  is also regarded as a random variable in $\mathcal{L}^2_{b,\theta}$. Assuming the foreground template is accurate ($\mathcal{A}=1$), the expectations are denoted as follows,
\beq
\llangle \vec{C}^{p, x}_b \rrangle = \vec{\mathbb{C}}^{p, x}_b, ~~\llangle \theta  \rrangle = \bar{\theta}, \llangle  \sum_x( \vec{A}^{p, T}_{x}\vec{C}^{p, x}_b ) \rrangle =  \sum_x( \vec{A}^{p, T}_{x}\vec{\mathbb{C}}^{p, x}_b ),\llangle \sum_x \vec{A}^{p, T}_{x}(\vec{C}^{p, x}_b- \vec{\mathbb{C}}^{p, x}_b ) \rrangle = 0
\eeq
notably because $\vec{C}^{p, x}_b$ is a function of polarization rotation angle parameters, we have
\beq
\llangle  \sum_x( \vec{A}^{p, T}_{x, \theta}\vec{C}^{p, x}_b ) \rrangle \neq  0
\eeq
we decompose  
$\vec{A}^{p, T}_{x,\theta}\vec{C}^{p, x}_b$
as follows to facilitate the computation of $\langle \mathcal{L}_{b,\theta}\rangle$,
\bea\label{eq:ul_derv_decomp}
\sum_x (\vec{A}^{p, T}_{x,\theta}\vec{C}^{p, x}_b) &=& \sum_x \left[\vec{A}^{p, T}_{x,\theta}\left(\vec{C}^{p, x}_b -\vec{\mathbb{C}}^{p, x}_b\right)+ \vec{A}^{p, T}_{x,\theta}\vec{\mathbb{C}}^{p, x}_b \right]
\eea

It is straightforward to verify that $\langle\mathcal{L}^2_{b,\theta}\rangle=0$. The term $\mathcal{L}^1_{b,\theta}$ contains no random variables. By substituting Eq.~(\ref{eq:ul_derv_decomp}) into the $\mathcal{L}^0_{b,\theta}$, taking expectation, and simplifying, we obtain,
\beq
\llangle \mathcal{L}^0_{b,\theta} \rrangle = \sum_{pq} \left( D^L_{b:pq,\theta}\Xi^{-1}_{b:pq} +\frac{1}{2}  \Xi_{b:pq} \Xi^{-1}_{b:pq, \theta}\right) = \frac{1}{2}   \sum_{pq} \left( \Xi_{b:pq,\theta}\Xi^{-1}_{b:pq} +\Xi_{b:pq} \Xi^{-1}_{b:pq, \theta}\right) = 0, 
\eeq
where the simplification makes use of the permutation symmetry of $p, q$ indices, as well as the matrix identity $\Xi^{-1}_{b, \theta} = -\Xi^{-1}_b\Xi_{b, \theta}\Xi^{-1}_b$. Thus we draw the conclusion $\langle\mathcal{L}_{b,\theta}\rangle=\mathcal{L}^1_{b,\theta}\neq0$.

We then follow the original definition of Fisher matrix, whose matrix element is identical to the product of two first-order of derivatives, to derive the analytical Fisher matrix expression,
\beq\label{eq:lhd_derv_prod_exp0}
\llangle \mathcal{L}_\theta \mathcal{L}_\phi \rrangle =\llangle\left[ \sum_{b_1} \left( \mathcal{L}^0_{b_1, \theta} +  \mathcal{L}^1_{b_1, \theta} \right) + \mathcal{L}^2_{,\theta}\right] \left[ \sum_{b_2} \left( \mathcal{L}^0_{b_2, \phi} +  \mathcal{L}^1_{b_2, \phi} \right) + \mathcal{L}^2_{,\phi}\right] \rrangle, 
\eeq

Assuming no correlation between the prior part and CMB power spectrum, the expectation of the cross-product between derivatives of $\mathcal{L}^2$
and those of the other two components vanishes. In addition, since $\langle \mathcal{L}^0_{,\theta} \rangle=0$, the expectation value of the cross-product between the derivatives of $\mathcal{L}^0$ and $\mathcal{L}^1$
also vanishes. Eq.~(\ref{eq:lhd_derv_prod_exp0}) therefore reduces to
\bea\label{eq:lhd_derv_exp}
\llangle \mathcal{L}_\theta \mathcal{L}_\phi \rrangle &=&\llangle\left[ \sum_{b_1} \left( \mathcal{L}^0_{b_1, \theta} +  \mathcal{L}^1_{b_1, \theta} \right) + \mathcal{L}^2_{,\theta}\right] \left[ \sum_{b_2} \left( \mathcal{L}^0_{b_2, \phi} +  \mathcal{L}^1_{b_2, \phi} \right) + \mathcal{L}^2_{,\phi}\right] \rrangle\nn\\
&=& \sum_{b_1} \llangle \mathcal{L}^0_{b_1, \theta}  \mathcal{L}^0_{b_1, \phi} \rrangle + \sum_{b_1 \neq b_2} \llangle \mathcal{L}^0_{b_1, \theta} \rrangle \llangle \mathcal{L}^0_{b_1, \phi} \rrangle + \sum_{b_1}  \mathcal{L}^1_{b_1, \theta} \sum_{b_2} \mathcal{L}^1_{b_2, \phi} + \llangle \mathcal{L}^2_{,\theta} \mathcal{L}^1_{b_2, \phi}\rrangle \nn\\
&=&  \sum_{b} \llangle \mathcal{L}^0_{b, \theta}  \mathcal{L}^0_{b, \phi} \rrangle  + \frac{1}{4} \left(\sum_{b}   \mathrm{Tr}( \Xi^{-1}_b\Xi_{b,\theta}) \right) \left(\sum_{b}    \mathrm{Tr}( \Xi^{-1}_b\Xi_{b,\phi}) \right) + \frac{\Delta_{\theta\phi}}{(\sigma_\theta^{cali})^2},
\eea
where we neglect the correlation of CMB power spectra between different bins. The definition of $\Delta_{\theta\phi}$ is provided in Eq.~(\ref{eq:fisher_matrix_expression}).

The main technical derivation is contained in the first part of Eq.~(\ref{eq:lhd_derv_exp}). By substituting Eq.~(\ref{eq:ul_derv_decomp}) into Eq.~(\ref{eq:lhd_derv_0}), and temporarily suppressing the index $b$ for clarity, we obtain:
\bea\label{eq:lhd_derv_00exp}
&&\llangle \mathcal{L}^0_{, \theta}  \mathcal{L}^0_{, \phi} \rrangle \nn\\
&=& \Big\langle\sum_{pq} \sum_x \left[\vec{A}^{p, T}_{x,\theta}\left(\vec{C}^{p, x}_b -\vec{\mathbb{C}}^{p, x}_b\right)  \Xi^{-1}_{pq} + \vec{A}^{p, T}_{x,\theta}\vec{\mathbb{C}}^{p, x}_b \Xi^{-1}_{pq} +\frac{1}{2}\vec{A}^{p, T}_{x} \vec{C}^{p, x}_b \Xi^{-1}_{pq, \theta}\right]\sum_y \left(\vec{A}^{q, T}_{y} \vec{C}^{q, y}_b\right)\nn\\
&&\times  \sum_{st} \sum_z \left[\vec{A}^{s, T}_{z,\phi}\left(\vec{C}^{s, z}_b -\vec{\mathbb{C}}^{s, z}_b\right)  \Xi^{-1}_{st} + \vec{A}^{s, T}_{z,\phi}\vec{\mathbb{C}}^{s, z}_b \Xi^{-1}_{st} +\frac{1}{2} \vec{A}^{s, T}_{z} \vec{C}^{s, z}_b  \Xi^{-1}_{st, \phi}\right]\sum_w \left(\vec{A}^{t, T}_{w} \vec{C}^{t, w}_b\right) \Big\rangle \nn\\
&=& \llangle \sum_{pq} (p^{(1)} + p^{(2)} + p^{(3)})q \sum_{st} (s^{(1)} + s^{(2)} + s^{(3)})t\rrangle 
\eea
in last step, we use indices as shorthand for their original expressions. According to Isserlis' theorem, the expectation of the product of an odd number of zero-mean Gaussian random variables vanishes, therefore,
\bea
\llangle \mathcal{L}^0_{b, \theta}  \mathcal{L}^0_{b, \phi} \rrangle 
&=& \llangle  (p^{(1)} +  p^{(3)})q  (s^{(1)}  + s^{(3)})t + p^{(2)}qs^{(2)}t\rrangle \\
&=& \sum_{pq}\sum_{st}  \llangle \left( p^{(1)}qs^{(1)}t + p^{(1)}qs^{(3)}t + p^{(3)}qs^{(1)}t + p^{(3)}qs^{(3)}t + p^{(2)}qs^{(2)}t\right) \rrangle\label{eq:lhd_derv_00prod_exp},\nn\\
\eea

Let us now simplify the five terms in the second line of Eq.~(\ref{eq:lhd_derv_00prod_exp}). For the first term, due to Isserlis' theorem,
\bea\label{eq:lhd_derv_00exp_1}
&&\sum_{pq}\sum_{st}  \llangle  p^{(1)}qs^{(1)}t\rrangle = \sum_{pq}\sum_{st} \left[ \llangle  p^{(1)}q\rrangle \llangle  s^{(1)}t\rrangle  + \llangle  p^{(1)}s^{(1)}\rrangle \llangle  qt\rrangle  +\llangle  p^{(1)}t\rrangle \llangle  qs^{(1)}\rrangle  \right]\nn\\
&=& \sum_{pq}\sum_{st} \left[ D^L_{pq, \theta} \Xi^{-1}_{pq} D^L_{st, \phi} \Xi^{-1}_{st} +   \sum_{xz}\vec{A}^{p, T}_{x,\theta}Q^{ps, x*z} \vec{A}^{s}_{z,\phi}\Xi^{-1}_{pq} \Xi^{-1}_{st} \Xi_{qt} +  D^L_{pt, \theta} \Xi^{-1}_{pq} D^L_{sq, \phi} \Xi^{-1}_{st} \right]\nn\\
&=&  \text{Tr}[ \Xi^{-1}D^L_{,\theta}]\text{Tr}[ \Xi^{-1}D^L_{,\phi}] + \frac{1}{2}\text{Tr}[\Xi^{-1}\Xi \Xi^{-1} \kappa_{\theta\phi} ]] + \text{Tr}[ \Xi^{-1}D^L_{,\phi} \Xi^{-1} D^L_{,\theta} ]  \nn\\
&=& \frac{1}{4}\text{Tr}[ \Xi^{-1}\Xi_\theta]\text{Tr}[ \Xi^{-1}\Xi_\phi]  + \frac{1}{2}\text{Tr}[\Xi^{-1} \kappa_{\theta\phi} ] + \text{Tr}[ \Xi^{-1}D^L_{,\phi} \Xi^{-1} D^L_{,\theta} ],
\eea
the last two steps utilize the permutation symmetry of the summation indices and the properties of the trace of a matrix, i.e.,
\bea
\mathrm{Tr}[ \Xi^{-1}D^L_{,\theta}] &=& \mathrm{Tr}[ (\Xi^{-1}D^L_{,\theta})^T] = \mathrm{Tr}[ \Xi^{-1}D^R_{,\theta}] = \frac{1}{2} \mathrm{Tr}[ \Xi^{-1}\Xi_{,\theta}], \nn\\
\kappa_{ps, \theta\phi} &=&  \sum_{xz}\left[ \vec{A}^{p, T}_{x,\theta}Q^{ps, x*z}\vec{A}^{s}_{z, \phi} + \vec{A}^{p, T}_{x,\phi}Q^{ps, x*z}\vec{A}^{s}_{z, \theta}\right], \kappa_{ps, \theta\phi} = \kappa_{sp, \theta\phi}
\eea

For the second term in Eq.~(\ref{eq:lhd_derv_00prod_exp}),similarly we have, 
\bea\label{eq:lhd_derv_00exp_2}
\sum_{pq}\sum_{st}  \llangle  p^{(1)}qs^{(3)}t\rrangle &=&  \sum_{pq}\sum_{st} \left[ \llangle  p^{(1)}q\rrangle \llangle  s^{(3)}t\rrangle  + \llangle  p^{(1)}s^{(3)}\rrangle \llangle  qt\rrangle  +\llangle  p^{(1)}t\rrangle \llangle  qs^{(3)}\rrangle  \right]\nn\\
&=&  \frac{1}{2} \sum_{pqst}\Xi^{-1}_{pq}  \Xi^{-1}_{st, \phi}\Big[  D^L_{pq,\theta}\Xi_{st} +  D^L_{ps,\theta}\Xi_{qt} + D^L_{pt,\theta}\Xi_{qs} \Big]\nn\\
&=& \frac{1}{4} \text{Tr}[\Xi^{-1}\Xi_{, \theta}] \text{Tr}[\Xi^{-1}_{, \phi} \Xi] + \frac{1}{2}\text{Tr}[\Xi^{-1}D^L_{, \theta}  \Xi^{-1}_{, \phi} \Xi] +  \frac{1}{2}\text{Tr}[\Xi^{-1} D^L_{, \theta} \Xi^{-1}_{, \phi}\Xi  ]\nn\\
&=&  -\frac{1}{4} \text{Tr}[\Xi^{-1}\Xi_{, \theta}] \text{Tr}[\Xi^{-1}\Xi_{, \phi} ] - \frac{1}{2} \text{Tr}[ \Xi^{-1} \Xi_{, \theta}\Xi^{-1} \Xi_{, \phi} ] 
\eea

By performing the index interchange $\theta \leftrightarrow \phi, p \leftrightarrow s, q \leftrightarrow t$, the second term in Eq.~(\ref{eq:lhd_derv_00prod_exp}) is converted into the third term. Hence, applying the same interchange $\theta \leftrightarrow \phi$ to the last line of  
Eq.~(\ref{eq:lhd_derv_00exp_2}) yields
\beq\label{eq:lhd_derv_00exp_3}
\sum_{pq}\sum_{st}  \llangle  p^{(3)}qs^{(1)}t\rrangle=  -\frac{1}{4} \text{Tr}[\Xi^{-1}\Xi_{, \theta}] \text{Tr}[\Xi^{-1}\Xi_{, \phi} ] - \frac{1}{2} \text{Tr}[ \Xi^{-1} \Xi_{, \theta}\Xi^{-1} \Xi_{, \phi} ],
\eeq

Repeat the simplification procedure to the fourth term in Eq.~(\ref{eq:lhd_derv_00prod_exp}), we obtain
\bea\label{eq:lhd_derv_00exp_4}
\sum_{pq}\sum_{st}  \llangle  p^{(3)}qs^{(3)}t\rrangle &=&  \sum_{pq}\sum_{st} \left[ \llangle  p^{(3)}q\rrangle \llangle  s^{(3)}t\rrangle  + \llangle  p^{(3)}s^{(3)}\rrangle \llangle  qt\rrangle  +\llangle  p^{(3)}t\rrangle \llangle  qs^{(3)}\rrangle  \right]\nn\\
&=& \frac{1}{4}\sum_{pqst} \Xi^{-1}_{pq, \theta}\Xi^{-1}_{st, \phi}\Big[ \Xi_{pq}\Xi_{st} + \Xi_{ps}\Xi_{qt} +\Xi_{pt}\Xi_{qs}\Big] \nn\\
&=&\frac{1}{4}\big(\mathrm{Tr}[\Xi \Xi^{-1}_{,\theta}]\mathrm{Tr}[\Xi\Xi^{-1}_{,\phi}]  +2 \mathrm{Tr}[\Xi^{-1}_{,\theta}\Xi\Xi^{-1}_{,\phi}\Xi] \big)\nn\\
&=& \frac{1}{4} \text{Tr}[\Xi^{-1}\Xi_{, \theta}] \text{Tr}[\Xi^{-1}\Xi_{, \phi} ] + \frac{1}{2} \text{Tr}[ \Xi^{-1} \Xi_{, \theta}\Xi^{-1} \Xi_{, \phi} ] 
\eea

The fifth term in Eq.~(\ref{eq:lhd_derv_00prod_exp}) can be simplified as follows,
\bea\label{eq:lhd_derv_00exp_5}
\sum_{pq}\sum_{st}  \llangle  p^{(2)}qs^{(2)}t\rrangle &=&  \sum_{pq}\sum_{st} p^{(2)}s^{(2)} \llangle  qt \rrangle  \nn\\
&=&  \sum_{pq} \sum_{st} \Xi^{-1}_{pq}\Xi^{-1}_{st}\sum_{x}\left( \vec{A}^{p, T}_{x,\theta} \vec{\mathbb{C}}^{p,x}\right) \sum_z \left( \vec{A}^{s, T}_{z,\phi}\vec{\mathbb{C}}^{s,z}\right)\Xi_{qt}\nn\\
&\overset{p \leftrightarrow s, q \leftrightarrow t}{=}&   \frac{1}{2}\text{Tr}[ \Xi^{-1}\lambda_{\theta\phi}], 
\eea
where we define the matrix symbol,
\beq
\lambda_{ps, \theta\phi} = \sum_{x}\left( \vec{A}^{p, T}_{x,\theta} \vec{\mathbb{C}}^{p,x}\right) \sum_z \left( \vec{A}^{s, T}_{z,\phi}\vec{\mathbb{C}}^{s,z}\right)  + (s \leftrightarrow p), \lambda_{ps, \theta\phi} = \lambda_{sp, \theta\phi}.
\eeq

Substituting the final expressions of Eqs.~(\ref{eq:lhd_derv_00exp_1},\ref{eq:lhd_derv_00exp_2},\ref{eq:lhd_derv_00exp_3},\ref{eq:lhd_derv_00exp_4},\ref{eq:lhd_derv_00exp_5}) into Eq.~(\ref{eq:lhd_derv_00exp}), performing matrix simplification, and using the obtained result to update Eq.~(\ref{eq:lhd_derv_exp}), restoring the hidden index $b$, we obtain the final expression for Fisher matrix,
\begin{eqnarray}\label{eq:fisher_matrix_form1_final}
\llangle \mathcal{L}_\theta \mathcal{L}_\phi \rrangle 
&=&  \sum_{b}\left[ \frac{1}{2}\text{Tr}[\Xi^{-1}_b (\kappa_{b\theta\phi}+\lambda_{b\theta\phi}) ] + \text{Tr}[ \Xi^{-1}_b D^L_{b,\phi} \Xi^{-1} D^L_{b,\theta} ]  - \frac{1}{2} \text{Tr}[ \Xi^{-1}_b \Xi_{b, \theta}\Xi^{-1}_b \Xi_{b, \phi} ]\right] \nn\\
&& + \frac{1}{4} \left(\sum_{b}   \mathrm{Tr}( \Xi^{-1}_b\Xi_{b,\theta}) \right) \left(\sum_{b}    \mathrm{Tr}( \Xi^{-1}_b\Xi_{b,\phi}) \right) + \frac{\Delta_{\theta\phi}}{(\sigma_\theta^{cali})^2},
\end{eqnarray}

Because the covariance matrix $\Xi$ depends on the parameter $\alpha$ and $\langle\mathcal{L}_{b,\theta}\rangle \neq0$, the equivalence between Fisher matrix defined from the expectation value of the product of first-order derivatives and that defined from the expectation value of the second-order derivatives is violated~\cite{2002Statistical}. For completeness, we therefore also derive Fisher matrix expression based on second-order derivatives. By further differentiating the first-order derivatives in Eq.~(\ref{eq:lhd_derv_0}), we obtain
\bea\label{eq:lhd_derv_2}
\mathcal{L}^0_{b,\theta\phi} &=& \sum_{pq} \left[\sum_x \left( \vec{A}^{p, T}_{x,\theta}\vec{C}^{p, x}_b \right)\Xi^{-1}_{b:pq} +\frac{1}{2} \sum_x \left( \vec{A}^{p, T}_x \vec{C}^{p, x}_b \right) \Xi^{-1}_{b:pq, \theta}\right]\sum_y \left(\vec{A}^{q, T}_{y,\phi}\vec{C}^{q, y}_b \right)\nn\\
&& + \sum_{pq}\sum_y \left(\vec{A}^{q, T}_y\vec{C}^{q, y}_b \right) \Big[ \sum_x \left( \vec{A}^{p, T}_{x,\theta\phi}\vec{C}^{p, x}_b \right)\Xi^{-1}_{b:pq}  + \sum_x\left( \vec{A}^{p, T}_{x,\theta}\vec{C}^{p, x}_b \right)\Xi^{-1}_{b:pq, \phi}\nn\\
&&+\frac{1}{2} \sum_x \left( \vec{A}^{p, T}_{x, \phi}\vec{C}^{p, x}_b\right) \Xi^{-1}_{b:pq, \theta} + \frac{1}{2}  \sum_x\left( \vec{A}^{p, T}_x\vec{C}^{p, x}_b\right) \Xi^{-1}_{b:pq, \theta\phi}\Big],\nn\\
\mathcal{L}^1_{b,\theta\phi} &=& \frac{1}{2} \mathrm{Tr}( \Xi^{-1}_b\Xi_{b,\theta\phi} - \Xi^{-1}_b\Xi_{b,\theta}\Xi^{-1}_b\Xi_{b,\phi}), ~~\mathcal{L}^2_{,\theta\phi} = \frac{\Delta_{\theta\phi}}{(\sigma_\theta^{cali})^2},
\eea
Recall that the second-order derivatives of the covariance matrix can be expanded as
\beq
\Xi^{-1}_{pq, \theta\phi}= \left[\Xi^{-1}\left( \Xi_{,\phi}\Xi^{-1}\Xi_{,\theta}  +   \Xi_{,\theta}\Xi^{-1}\Xi_{,\phi}  - \Xi_{,\theta\phi}\right) \Xi^{-1} \right]_{pq}
\eeq
then the first term in Eq.~(\ref{eq:lhd_derv_2}) can be simplified as
\bea
\llangle \mathcal{L}^0_{b,\theta\phi} \rrangle   &=&  \sum_{pq} \sum_x \left[ \vec{A}^{p, T}_{x,\theta} \left(\vec{C}^{p, x}_b- \vec{\mathbb{C}}^{p, x}_b \right)\Xi^{-1}_{b:pq}  + \vec{A}^{p, T}_{x,\theta}  \vec{\mathbb{C}}^{p, x}_b \Xi^{-1}_{b:pq} +\frac{1}{2}   \vec{A}^{p, T}_x \vec{C}^{p, x}_b  \Xi^{-1}_{b:pq, \theta}\right] \nn\\
&&\times \sum_y \vec{A}^{q, T}_{y,\phi} \left(\vec{C}^{q, y}_b - \vec{\mathbb{C}}^{q, y}_b +\vec{\mathbb{C}}^{q, y}_b \right)+ \sum_{pq}  \sum_x  \Big[\vec{A}^{p, T}_{x,\theta\phi} \left( \vec{C}^{p, x}_b- \vec{\mathbb{C}}^{p, x}_b \right)\Xi^{-1}_{b:pq} + \vec{A}^{p, T}_{x,\theta\phi}  \vec{\mathbb{C}}^{p, x}_b \Xi^{-1}_{b:pq}   \nn\\
&&  +  \vec{A}^{p, T}_{x,\theta} \left( \vec{C}^{p, x}_b -\vec{\mathbb{C}}^{p, x}_b \right)\Xi^{-1}_{b:pq, \phi} +\vec{A}^{p, T}_{x,\theta} \vec{\mathbb{C}}^{p, x}_b  \Xi^{-1}_{b:pq, \phi}+\frac{1}{2}\vec{A}^{p, T}_{x, \phi} \left( \vec{C}^{p, x}_b - \vec{\mathbb{C}}^{p, x}_b\right) \Xi^{-1}_{b:pq, \theta}  \nn\\
&& +\frac{1}{2}\vec{A}^{p, T}_{x, \phi}\vec{\mathbb{C}}^{p, x}_b \Xi^{-1}_{b:pq, \theta}+ \frac{1}{2} \left( \vec{A}^{p, T}_x\vec{C}^{p, x}_b\right) \Xi^{-1}_{b:pq, \theta\phi}\Big] \sum_y \left(\vec{A}^{q, T}_y\vec{C}^{q, y}_b \right) \nn\\ 
&= &  \sum_{pq}  \sum_{xy} \Big[ \vec{A}^{p, T}_{x,\theta} Q^{pq, x*y} \vec{A}^{q}_{y,\phi}\Xi^{-1}_{b:pq} +\vec{A}^{p, T}_{x,\theta}  \vec{\mathbb{C}}^{p, x}_b \Xi^{-1}_{b:pq} \vec{A}^{q, T}_{y,\phi} \vec{\mathbb{C}}^{q, y}_b +\frac{1}{2}   \vec{A}^{p, T}_x Q^{pq, x*y} \vec{A}^{q}_{y,\phi} \Xi^{-1}_{b:pq, \theta}  \nn\\
&&+ \vec{A}^{p, T}_{x,\theta\phi}  Q^{pq, x*y}  \vec{A}^{q}_y \Xi^{-1}_{b:pq} + \vec{A}^{p, T}_{x,\theta}  Q^{pq, x*y} \vec{A}^{q}_y \Xi^{-1}_{b:pq, \phi} + \frac{1}{2}\vec{A}^{p, T}_{x, \phi} Q^{pq, x*y} \vec{A}^{q}_y  \Xi^{-1}_{b:pq, \theta} \nn\\
&& +   \frac{1}{2}   \vec{A}^{p, T}_x  Q^{pq, x*y} \vec{A}^{q}_y \Xi^{-1}_{b:pq, \theta\phi} \Big]\nn\\
&=& \frac{1}{2}\text{Tr}[\Xi^{-1}_b \lambda_{b\theta\phi} ],
\eea

Following a procedure similar to that described above, we obtain a simplified expression for the Fisher matrix defined in terms of second-order derivatives, i.e.,
\bea\label{eq:fisher_matrix_form2}
\llangle \mathcal{L}_{\theta\phi} \rrangle &=& \sum_b  \frac{1}{2}\left[ \text{Tr}(\Xi^{-1}_b \lambda_{b\theta\phi} ) + \mathrm{Tr}( \Xi^{-1}_b\Xi_{b,\theta\phi} - \Xi^{-1}_b\Xi_{b,\theta}\Xi^{-1}_b\Xi_{b,\phi})\right] +  \frac{\Delta_{\theta\phi}}{(\sigma_\theta^{cali})^2}
\eea
where the second-order derivatives of covariance matrix $\Xi$ is calculated as, 
\bea
\Xi_{, \theta\phi} &=&\sum_{xy} \left[ \vec{A}^{p, T}_{x,\theta\phi} Q^{pq, x*y} \vec{A}^{q}_y + \vec{A}^{p, T}_{x,\theta} Q^{pq, x*y} \vec{A}^{q}_{y,\phi} + \vec{A}^{p, T}_{x,\phi} Q^{pq,x*y} \vec{A}^{q}_{y,\theta} + \vec{A}^{p, T}_x Q^{pq, x*y} \vec{A}^{q}_{y,\theta\phi}\right],\nn\\
\vec{A}^{p, (0)}_{o, \alpha^t\alpha^s} &=&  \frac{16}{\left[\cos(4\alpha^i)+\cos(4\alpha^j)  \right]^2}\Big[  \sin(4\alpha^i)\cos(4\alpha^j)(\delta_{is}\delta_{jt} - \delta_{it}\delta_{js})  +  \cos(4\alpha^i)\sin(4\alpha^j)(\delta_{js}\delta_{jt} - \delta_{is}\delta_{it})  \Big]\nn\\
&& -\frac{32}{\left[\cos(4\alpha^i)+\cos(4\alpha^j)  \right]^3}\big [\left(1+\cos(4\alpha^i)\cos(4\alpha^j)\right)\delta_{jt} +\sin(4\alpha^i)\sin(4\alpha^j)\delta_{it}\big]\nn\\
&&\times \big[\delta_{is}\sin(4\alpha^i) +\delta_{js}\sin(4\alpha^j)\big],\nn\\
\vec{A}^{p, (1)}_{o, \alpha^t\alpha^s} &=& -\vec{A}^{p, (0)}_{o, \alpha^t\alpha^s}(i \leftrightarrow j), ~~~\vec{A}^{p, (3)}_{, \alpha^t\alpha^s}  = 0, \nn\\
\vec{A}^{p, (0)}_{c, \alpha^t\beta} &=&  \frac{4\cos(4\beta)\tan2(\alpha^{i}+\alpha^{j})}{\cos2(\alpha^{i}+\alpha^{j})}(\delta_{p_it}+\delta_{p_jt})
,\vec{A}^{p, (1)}_{c, \alpha^t\beta}= -\vec{A}^{p, (0)}_{c, \alpha^t\beta} \nn\\
\vec{A}^{p, (0)}_{c, \alpha^t\alpha^s} &=&  \frac{\sin(4\beta)\left(3-\cos 4(\alpha^{i}+\alpha^{j}) \right)}{\cos^3 2(\alpha^{i}+\alpha^{j})}(\delta_{p_it}+\delta_{p_jt})(\delta_{p_is}+\delta_{p_js})
, \vec{A}^{p, (1)}_{c, \alpha^t\alpha^s} = -\vec{A}^{p, (0)}_{c, \alpha^t\alpha^s}\nn\\
\vec{A}^{p, (0)}_{f, \alpha^t\alpha^s} &=& \mathcal{A}\Big[\frac{\left(3-\cos 4(\alpha^{i}+\mathbf{1}\cdot\alpha^{j}) \right)}{\cos^3 2(\alpha^{i}+\mathbf{1}\cdot\alpha^{j})}(\delta_{p_it}+\mathbf{1}\cdot\delta_{p_jt})(\delta_{p_is}+\mathbf{1}\cdot\delta_{p_js}) +(\mathbf{1} \rightarrow -\mathbf{1}) \Big],\nn\\
\vec{A}^{p, (1)}_{f, \alpha^t \alpha^s}&=&\mathcal{A}\Big[\frac{\left(3-\cos 4(\alpha^{i}+\mathbf{1}\cdot\alpha^{j}) \right)}{\cos^3 2(\alpha^{i}+\mathbf{1}\cdot\alpha^{j})}(\delta_{p_it}+\mathbf{1}\cdot\delta_{p_jt})(\delta_{p_is}+\mathbf{1}\cdot\delta_{p_js}) -(\mathbf{1} \rightarrow -\mathbf{1}) \Big],\nn\\
\vec{A}^{p, (1)}_{f, \alpha^t \mathcal{A}} &=& \vec{A}^{p, (1)}_{f, \alpha^t }(\mathcal{A}=1)
\eea

\begin{figure}[htbp]
	\centering
	\includegraphics[width=0.9\columnwidth]{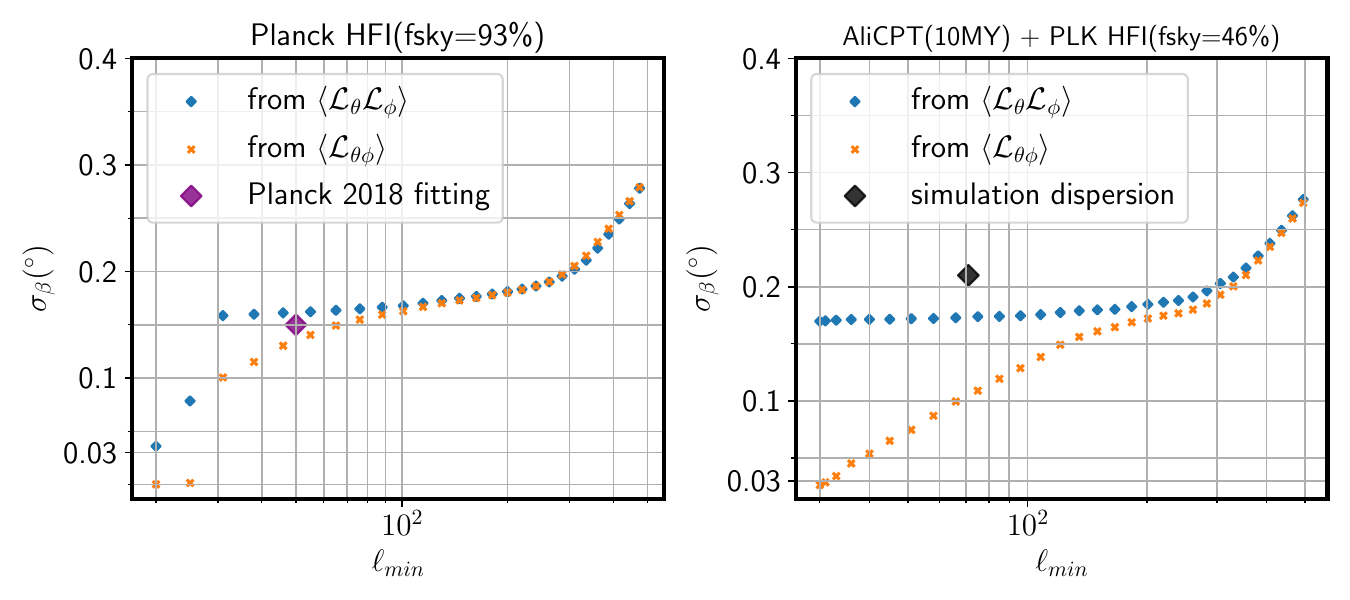}
	\caption{Comparison of forecasts for $\sigma_\beta$ as a function of $\ell_{min}$(with $\ell_{max}=1500$), obtained using two types of Fisher matrix, without any foreground template and calibration prior. The left panel shows the forecast based on the \textit{Planck} HFI over a 93\% sky fraction, as adopted in the Ref.~\cite{Minami:2020odp}, while the right panel presents the joint forecast from AliCPT-1(10 module-years of observation) and \textit{Planck} HFI over AliCPT-1’s 46\% sky coverage.}\label{fig:fisher_comparison}
\end{figure}

Figure.~\ref{fig:fisher_comparison} illustrate the different forecasts for $\sigma_\beta$ obtained from two types of Fisher matrix. For nearly full sky observation,  the two forecasts are consistent with the fitting result from Ref.~\cite{Minami:2020odp} while $\ell_{min}=51$. In contrast, for sky coverage of AliCPT-1, the forecast based on $\llangle \mathcal{L}_{,\theta\phi} \rrangle$ is systematically smaller than that based on  $\llangle \mathcal{L}_{,\theta} \mathcal{L}_{,\phi} \rrangle$. In particular, when $\ell_{min}<100$, $\llangle \mathcal{L}_{,\theta\phi} \rrangle$ significantly underestimates $\sigma_\beta$ compared to simulation dispersion value when $\ell_{min}=71$. This discrepancy motivates our choice of using  $\llangle \mathcal{L}_{,\theta} \mathcal{L}_{,\phi} \rrangle$ rather than $\llangle \mathcal{L}_{,\theta\phi} \rrangle$ for the forecasts presented in this paper.

\section{Simulation forecast for isotropic polarization rotation angle}
\label{append:simulation}

The Simulation constraints on the isotropic rotation angle $\beta$ presented in Figure.~\ref{fig:sig_beta_vs_noise} are derived from the statistical analysis of minimum likelihood estimations across 200 datasets.
These simulations combine the AliCPT-1 dataset (10 module-years) with \textit{Planck} HFI data. The detailed simulation procedure is described below.

\subsection{Data simulation}
\begin{figure}[htbp]
	\centering    
	\includegraphics[width=0.8\columnwidth]{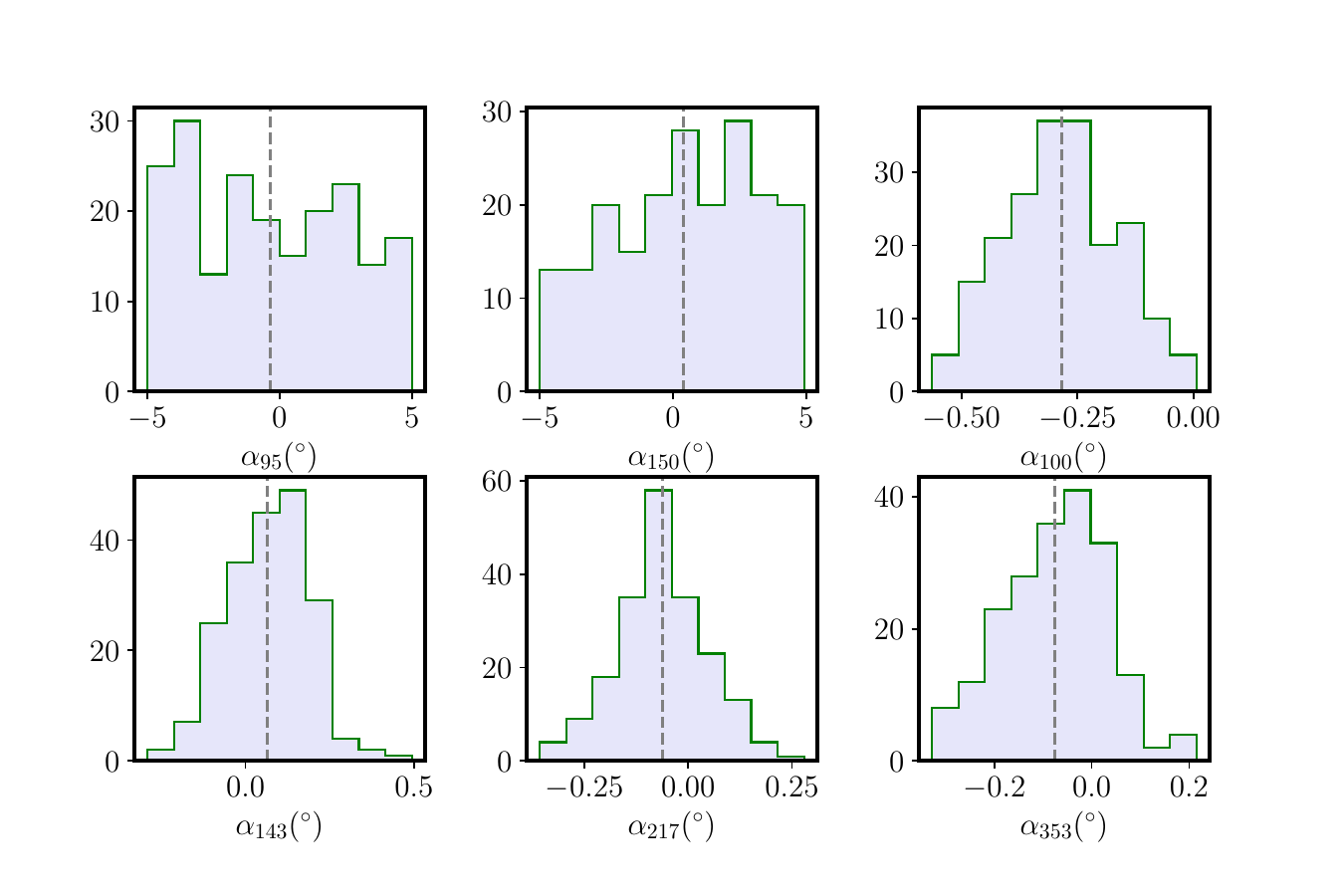}
	\caption{The input values for the 200 randomly generated polarization miscalibration angles.}\label{fig:input_rotation}
\end{figure}

\begin{figure}[htbp]
	\centering    
	\includegraphics[width=0.6\columnwidth]{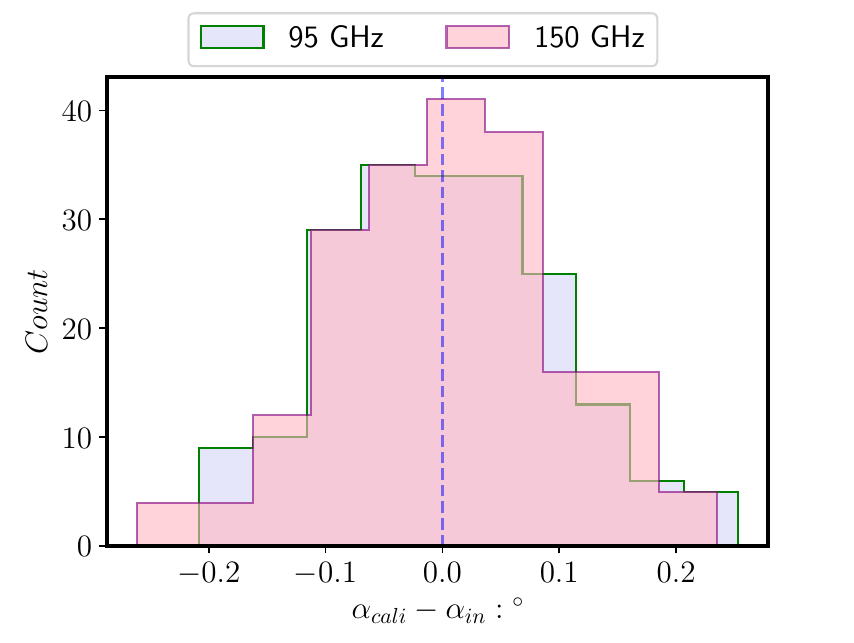}
	\caption{Difference between the calibrated polarization miscalibration angle value $\alpha_{\text{cali}}$ and the input true value $\alpha_{\text{in}}$.}\label{fig:priors}
\end{figure}

We generated 200 sets of CMB, foreground, and noise simulation maps with $nside=1024$ for all six bands of AliCPT-1 and Planck HFI. 
The simulation steps are as follows:
\begin{itemize}
	\item CMB map Simulation: 
	Using the best-fit $\Lambda$CDM cosmological parameters from \textit{Planck} 2018, we compute the theoretical angular power spectra $C_\ell$ with the \texttt{CAMB} code~\cite{Lewis:1999bs}, including scalar perturbations and lensing effects but excluding primordial tensors.
    These spectra are then used to generate 200 full-sky realizations of CMB polarization maps with \texttt{Healpy}.

	\item Foreground map Simulation: 
    {Foreground polarization maps are generated using the \texttt{PySM} medium-complexity model}.
    For the two AliCPT-1 bands, we convolve the foreground spectra with ideal top-hat bandpass functions, while for the \textit{Planck} HFI channels we use the measured instrumental bandpass responses.

	\item Noise Simulation:
	For AliCPT-1, {the noise is simulated using the noise levels that are scaled to correspond to 10 module-years of observation, based on the values presented in Table.~\ref{tab:bands_design}}.
    Gaussian random noise realizations are then generated.
    For \textit{Planck} HFI, we directly adopt the FFP10 noise simulation maps.
	
	\item Polarization Rotation Angle Simulation:
    We introduce randomness in polarization miscalibration angle $\alpha_{\mathrm{in},i}$.
    For the two AliCPT-1 bands, we draw $\alpha_{\mathrm{in},i}$ from a uniform distribution within $\pm5^\circ$, while for the four \textit{Planck} HFI bands, $\alpha_{\mathrm{in},i}$ are drawn from Gaussian distributions with means and variances taken from Table.~1 of Ref.~\cite{Minami:2020odp}.
    The Figure.~\ref{fig:input_rotation} shows the distribution of input miscalibration angle $\alpha_{\text{in}}$ across all realizations.
    The Chern-Simons rotation angle $\beta$ is fixed to $0.35^\circ$.
    
    \item Sky Map Rotation and Convolution: 
    Each simulated CMB map is rotated by $(\alpha_{\mathrm{in},i} + \beta)$, whereas the foreground maps are rotated only by $\alpha_i$.
    The CMB and foreground components are then co-added and convolved with a Gaussian beam corresponding to each frequency band.
 
	\item Map coaddition: 
    We add the rotated maps and the noise maps to obtain the observed maps.
    In total, 200 independent full-sky realizations are produced for subsequent analysis.
    
    \item Calibration:
    We consider an external calibration with an accuracy of $0.1^\circ$ for AliCPT-1 dual bands. 
    For each simulated dataset, representing an independent experimental realization, we draw the calibration center value $\alpha_{\text{cali}}$ from $\mathcal{N}(\alpha_{\text{in}}, 0.1\si{\degree})$.
    The Figure.~\ref{fig:priors} shows the distribution of the differences $\alpha_{\text{cali}} - \alpha_{\text{in}}$ across all realizations.
    In subsequent analysis, $\alpha_{\text{cali}}$ will be used as $\bar{\alpha_i}$ in the likelihood function Eq.~(\ref{eq:likelihood_eq}).
\end{itemize}

\subsection{Result}
We computed the polarization power spectrum from the simulated sky maps using NaMaster. 
A mask covering 46\% of the sky with $1\si{\degree}$ apodization was applied, and E and B mode purification was enabled to mitigate E-to-B leakage.
The spectrum was binned with a width of $\Delta \ell=20$. 
In evaluating the covariance matrix according to Eq.~(\ref{eq:binned_cov}), the theoretical spectrum was approximated by the observed power spectrum. 
We found that the covariance matrices $\Xi_{b}$ for the first three bins ($b=1,2,3$) were not positive definite. Therefore, these bins were excluded, and the analysis was started from the fourth bin, corresponding to $\ell_{\min}=71$, with $\ell_{\max}$ set to 1500.

\begin{figure}[htbp]
	\centering    
	\includegraphics[width=0.9\columnwidth]{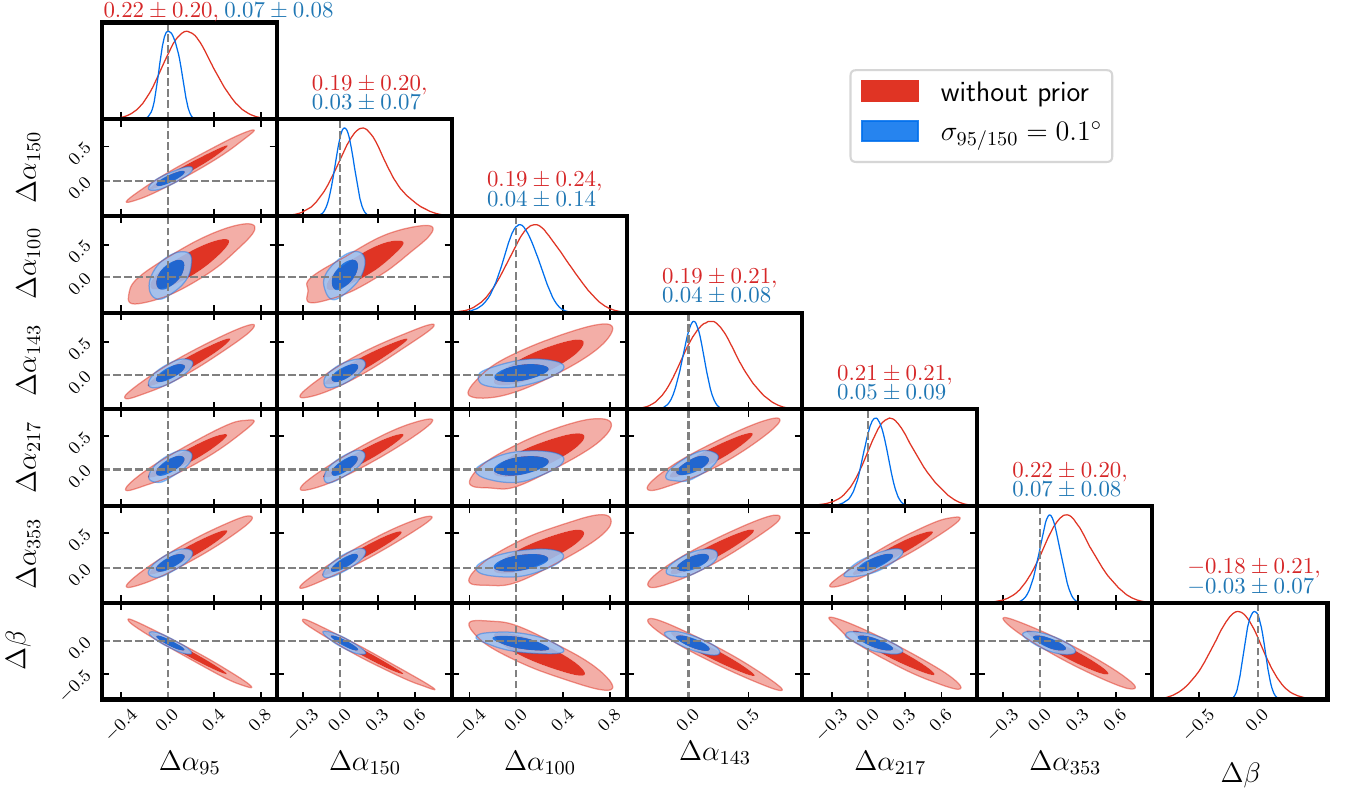}
	\caption{The sample distribution and statistical results obtained from 200 sets of best-fit values of polarization rotation angle parameters, where $\Delta \alpha= \alpha-\alpha_{in}$  and $\Delta \beta =\beta-\beta_{in}$. Foreground templates are not used in this analysis.}\label{fig:triangle_const_angle}
\end{figure}

We asses the effect of external calibration on the measurement uncertainty of $\beta$ by comparing two types of simulation analyses. The first includes a Gaussian prior on $\alpha$ to emulate the use of calibration, while the second excludes this prior, corresponding to the scenario without external calibration.
Figure.~\ref{fig:triangle_const_angle} displays the distribution of the minimum likelihood estimated values obtained from 200 simulated datasets.
With the prior included, the posterior mean of $\beta$ exhibits a markedly reduced offset from the fiducial value, and its uncertainty decreases from $0.21\si{\degree}$ to $0.07\si{\degree}$. In presence of a calibration prior, the miscalibration angle $\alpha$ is effectively anchored near its true value, thereby mitigating the propagation of errors into the estimate of $\beta$.

Notably, the current analysis applies the polarization angle prior only to the two AliCPT-1 bands. Including analogous priors for the \textit{Planck} HFI like channels would yield tighter constraints on the polarization angles and a further reduction in the uncertainties.

\bibliographystyle{unsrt}
\bibliography{reference}	
	
\end{document}